\documentclass[preprint,showpacs,preprintnumbers,amsmath,amssymb]{revtex4-1}
\usepackage{graphicx, latexsym, amssymb, amsmath, color, multirow, mathrsfs, booktabs, ifpdf}
\usepackage{dcolumn}
\usepackage{bm}

\def\phe6{$^6$He+$p$\ }
\def\he6pn{$p(^6$He,$^6$Li$^*)n$\ }

\def\pn{$(p,n)$\ }

\begin{document}
\title{Equation of state of the neutron star matter, and the nuclear symmetry energy}
\author{Doan Thi Loan$^{1}$}
\author{Ngo Hai Tan$^{1}$}
\author{Dao T. Khoa$^1$}\email{Corresponding author: khoa@vaec.gov.vn}
\author{Jerome Margueron$^2$}
\affiliation{$^1$Institute for Nuclear Science and Technique, VAEC \\ 179
Hoang Quoc Viet Road, Nghia Do, Hanoi, Vietnam. \\
 $^2$ Institut de Physique Nucl\'eaire, IN2P3-CNRS/ Universit\'e Paris-Sud,
 91406 Orsay, France.}

\date{\today}

\begin{abstract}
The nuclear mean-field potentials obtained in the Hartree-Fock method with different
choices of the in-medium nucleon-nucleon (NN) interaction have been used
to study the equation of state (EOS) of the neutron star (NS) matter. The EOS of the
uniform NS core has been calculated for the np$e\mu$ composition in the
$\beta$-equilibrium at zero temperature, using version Sly4 of the Skyrme
interaction as well as two density-dependent versions of the finite-range M3Y
interaction (CDM3Y$n$ and M3Y-P$n$), and versions D1S and D1N of the
Gogny interaction. Although the considered effective NN interactions were proven
to be quite realistic in numerous nuclear structure and/or reaction studies, they
give quite different behaviors of the symmetry energy of nuclear matter at
supranuclear densities that lead to the \emph{soft} and \emph{stiff} scenarios
discussed recently in the literature. Different EOS's of the NS core and the EOS
of the NS crust given by the compressible liquid drop model have been used
as input of the Tolman-Oppenheimer-Volkov equations to study how the nuclear
symmetry energy affects the model prediction of different NS properties, like the
cooling process as well as the gravitational mass, radius, and moment of inertia.

\end{abstract}
 \maketitle

\section{Introduction}
The determination of the equation of state (EOS) of asymmetric nuclear matter
(NM) has been the main object of numerous nuclear structure and reaction studies
involving unstable nuclei lying close to the neutron or proton driplines
 \cite{Ba08}. The knowledge about the EOS of asymmetric NM is vital  for any
model of neutron star \cite{Bet90,Su94,Su95,Dou01,Lat04,Kl06}, and the
nuclear mean-field potential is the most important input for the determination of
the nuclear EOS. Many microscopic studies of the EOS have been done based on
the nuclear mean field given by both nonrelativistic and relativistic nuclear
many-body approaches, using realistic two-body and three-body nucleon-nucleon
(NN) forces or interaction Lagrangians (see recent reviews \cite{Ba08,Bal07}).
These many-body studies have shown the important role played by the
Pauli blocking effects as well as higher-order NN correlations
at different NM densities. These medium effects are normally considered as
the physics origin of the density dependence that has been introduced into various
versions of the effective NN interactions used in the modern mean-field approaches.
Among them, very popular is the so-called M3Y interaction which was originally
constructed to reproduce the G-matrix elements of the Reid \cite{Be77} and
Paris \cite{An83} NN potentials in an oscillator basis.  Several realistic density
dependences have been added later on to the M3Y interactions
 \cite{Kho93,Kho95,Kho96,Kho97,Kho07,Basu08} to properly account for
the NM saturation properties as well as the ground-state structure of finite nuclei
\cite{Na02,Na03,Na08}. These density dependent versions of the M3Y interaction
have been used in the nonrelativistic Hartree-Fock (HF) studies of symmetric and
asymmetric NM. Some of them have been successfully used in the folding model
studies of the nucleon-nucleus and nucleus-nucleus scattering
\cite{Kho96,Kho97,Kho07,Kho05,Kho07r,Kho09}.

In attempt to find a realistic version of the effective NN interaction for consistent
use in the mean-field studies of NM and finite nuclei as well as in the nuclear
reaction calculations, we have performed recently a systematic HF study of NM
\cite{Tha09} using the CDM3Y$n$ interactions, which have been used mainly
in the folding model studies of the nuclear scattering \cite{Kho97,Kho07,Kho07r,
Kho09}, and the M3Y-P$n$ interactions carefully parametrized by Nakada
\cite{Na02,Na03,Na08} for use in the HF studies of nuclear structure.
For comparison, the same HF study has also been done with the D1S and D1N
versions of the Gogny interaction \cite{Be91,Ch08} and Sly4 version of the
Skyrme interaction \cite{Ch98}.
While these effective NN interactions give more or less the same description
of the saturation properties of the symmetric NM, the HF results for the asymmetric
NM \cite{Tha09} show that they are divided into two families, which
are associated with two different (\emph{soft} and \emph{stiff}) behaviors
of the NM symmetry energy at high nucleon densities. As a result, these two
families predict very different behaviors of the proton-to-neutron
ratio in the $\beta$-equilibrium that can imply two drastically different
mechanisms for the neutron star cooling (with or without the direct Urca
process) \cite{Lat91,Lat94,Pa04}.

As a further step in this direction, we try to find out in the present work how
such a difference in the NM symmetry energy can affect the EOS  of the
$\beta$-stable neutron star (NS) matter as well as the main NS properties like
the maximum mass, radius, central density and moment of inertia. For this
purpose, the Tolman-Oppenheimer-Volkov (TOV) equations have been solved
using different EOS's of the NS matter that are associated with the nuclear
mean-field potentials given by different in-medium NN interactions under study.
Given the complex, inhomogeneous  structure of the NS crust, it is a tremendous
task to develop a consistent structure model for the inner and outer NS crusts using
all versions of the in-medium NN interaction considered here. Therefore,
we have used the EOS of the NS crust given by the Compressible Liquid Drop
Model (CLDM) \cite{Dou01,Dou00} with the model parameters determined
by the SLy4 interaction \cite{Ch98}.
Different EOS's of the uniform NS core are then calculated for the np$e\mu$
composition in the $\beta$-equilibrium at zero temperature and extended
to the supranuclear densities, using the mean-field potentials given by different
density-dependent NN interactions. In this way, any difference found in the
solutions of the TOV equations is entirely due to the choice of the EOS of the
NS core, i.e., to the choice of the in-medium NN interaction. The main NS properties
obtained in each case are compared with the empirical data given by the recent
astronomical observation of neutron stars.

\section{Hartree-Fock calculation of asymmetric nuclear matter}

We recall here the main features of our HF study \cite{Tha09} of the uniform
(spin-saturated) NM at zero temperature that is characterized by
given values of the neutron and proton densities, $n_{\rm n}$ and $n_{\rm p}$, or
equivalently by the total density $n_{\rm b}=n_{\rm n}+n_{\rm p}$ (hereafter
referred to as the baryon density) and the neutron-proton
asymmetry $\delta=(n_{\rm n}-n_{\rm p})/(n_{\rm n}+n_{\rm p})$.
With the direct ($v_{\rm D}$) and exchange ($v_{\rm EX}$) parts of the
interaction determined from the singlet- and triplet-even (and odd) components
of the central NN force, the total energy density of the NM is determined as
\begin{equation}
\varepsilon=\varepsilon_{\rm kin}+{\frac{1}{ 2}}\sum_{k \sigma \tau}
\sum_{k'\sigma '\tau '} [\langle{\bm k}\sigma \tau, {\bm k}' \sigma' \tau'
 |v_{\rm D}|{\bm k}\sigma\tau, {\bm k}' \sigma' \tau' \rangle
+ \langle{\bm k}\sigma \tau, {\bm k}'\sigma' \tau' |v_{\rm EX}|
{\bm k}'\sigma \tau, {\bm k}\sigma' \tau' \rangle],
 \label{e2} \end{equation}
where $|{\bm k}\sigma \tau\rangle$ are the ordinary plane waves.
Dividing $\varepsilon$ over the total baryon number density $n_{\rm b}$,
we obtain the total NM energy per particle $E$ that can be expressed as
\begin{equation}
\frac{\varepsilon}{n_{\rm b}}\equiv E(n_{\rm b},\delta)=
 E(n_{\rm b},\delta=0)+ S(n_{\rm b})\delta^2+O(\delta^4)+... \label{e3}
\end{equation}
The NM pressure $P$ and incompressibility $K$ are then calculated as
\begin{equation} P(n_{\rm b},\delta)=n_{\rm b}^2
\frac{\partial E(n_{\rm b},\delta)}{\partial n_{\rm b}}; \
 K(n_{\rm b},\delta)=9n_{\rm b}^2\frac{\partial^2E(n_{\rm b},\delta)}
 {\partial n_{\rm b}^2}. \label {e4}
\end{equation}
The contribution of $O(\delta^4)$ and higher-order terms in Eq.~(\ref{e3}) was
proven to be quite small \cite{Kho96,Zuo99} and is often neglected in the
so-called \emph{parabolic} approximation, where the NM \emph{symmetry
energy} $S(n_{\rm b})$ equals the energy required per particle
to change the symmetric NM into the pure neutron matter. The value
of $S(n_{\rm b})$ at the symmetric NM saturation density, $n_0\approx
0.17$\ fm$^{-3}$, is known as the symmetry energy coefficient $J=S(n_0)$
that has been predicted by numerous many-body calculations to be around 30 MeV
\cite{Kho96,Zuo99,Bra85,Pea00}.

\begin{figure}[bht] 
\includegraphics[width=0.7\textwidth]{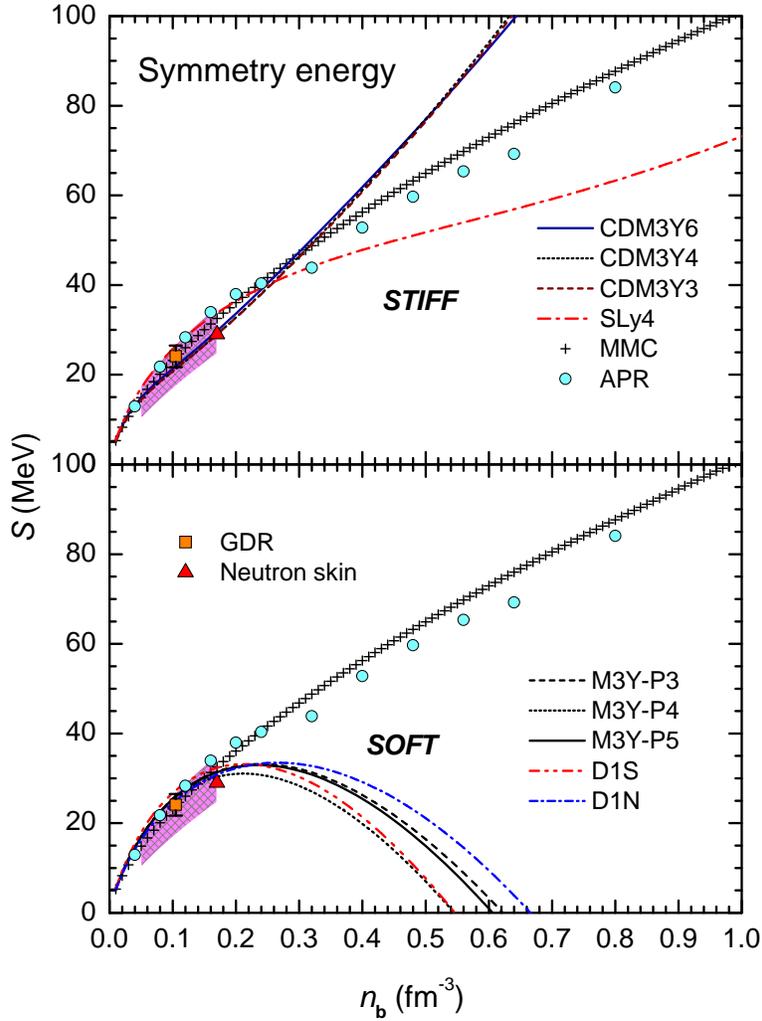}\vspace*{-1cm}
\caption{(Color online) HF results for the NM symmetry energies $S(n_{\rm b})$ given
by the density-dependent NN interactions under study. The shaded (magenta) region
marks the empirical boundaries deduced from the analysis of the isospin diffusion data
and double ratio of neutron and proton spectra data of HI collisions \cite{Ono03,Tsa09}.
The square and triangle are the constraints deduced from the consistent structure
studies of the GDR \cite{Tri08} and neutron skin \cite{Fur02}, respectively.
The circles and crosses are results of the ab-initio calculation by Akmal, Pandharipande
and Ravenhall (APR) \cite{Ak98} and microscopic Monte Carlo (MMC) calculation by
Gandolfi {\it et al.} \cite{Gan10}, respectively.} \label{f1}
\end{figure}

The knowledge about the density dependence of $S(n_{\rm b})$ is extremely
important for the construction of nuclear EOS  and it has been, therefore,
a longstanding goal of many nuclear structure and reaction studies.
The main method to probe $S(n_{\rm b})$ associated with a given in-medium NN
interaction is to test this interaction in the simulation of heavy-ion (HI)
collisions using transport and/or statistical models
\cite{Tsa09,Ono03,Ba08,Da02,Ch05,Ba09,She07,Sh07} or in the structure studies of
nuclei with large neutron excess \cite{Na08,Ch08,Ch98,Da03,Aru04,Tod05,Pie07,
Pie09,Cen09,Tri08,Br00,Fur02}. Based on the physics constraints implied by such studies,
extrapolation is often made to draw conclusion on the low- and high-density behavior
of $S(n_{\rm b})$. However, such conclusions still remain quite divergent in some
cases \cite{Tha09}. One of the most intriguing issues discussed recently in the
literature is whether the ``soft" or ``stiff" density dependence of the NM symmetry
energy is more realistic. These two scenarios are well illustrated in Fig.~\ref{f1}
where the HF results obtained with the two groups of the considered
in-medium NN interactions are plotted.
For comparison, we have also plotted in Fig.~\ref{f1} results of the ab-initio
variational chain-summation calculation using the A18+$\delta v$+UIX* version
of the Argonne NN potential by Akmal, Pandharipande and Ravenhall (APR)
\cite{Ak98} and recent microscopic Monte Carlo calculation of neutron star structure
by Gandolfi {\it et al.} \cite{Gan10}, using the Argonne AV6' potential added by an
empirical density dependence.

Around the saturation density $n_0$ of the symmetric NM all the models
predict the symmetry coefficient $S(n_0)=J\approx 29\pm 3$ MeV, in a reasonable
agreement with the empirical values deduced recently from  the structure studies of
of neutron skin  \cite{Br00,Fur02}. In the low-density region ($n_{\rm b}\approx
0.3 \sim 0.6\ n_0$) there exist empirical boundaries for the symmetry energy
deduced from the analysis of the isospin diffusion data and double ratio of neutron
and proton spectra data of HI collisions \cite{Ono03,Tsa09}, which enclose the
HF results given by both groups of the in-medium NN interactions. At the baryon
density $n_{\rm b}\approx 0.1$ fm$^{-3}$, all the HF results also agree quite
well with the empirical value deduced from a consistent structure study of the
isovector Giant Dipole Resonance (GDR) in heavy nuclei \cite{Tri08}.
So far there are no firm empirical constraints on the NM symmetry energy at
supranuclear densities, and the behavior of $S(n_{\rm b})$ at high densities
remains uncertain. The two different behaviors of $S(n_{\rm b})$ shown
in Fig.~\ref{f1} have been observed earlier \cite{Ba08,She07,Ba05,Sto03} and
often discussed in the literature as the \emph{Asy-stiff} (with symmetry energy
steadily increasing with density) and \emph{Asy-soft} (with symmetry energy
reaching saturation and then decreasing to negative values) behaviors.
\begin{figure}[bht] 
\includegraphics[width=0.7\textwidth]{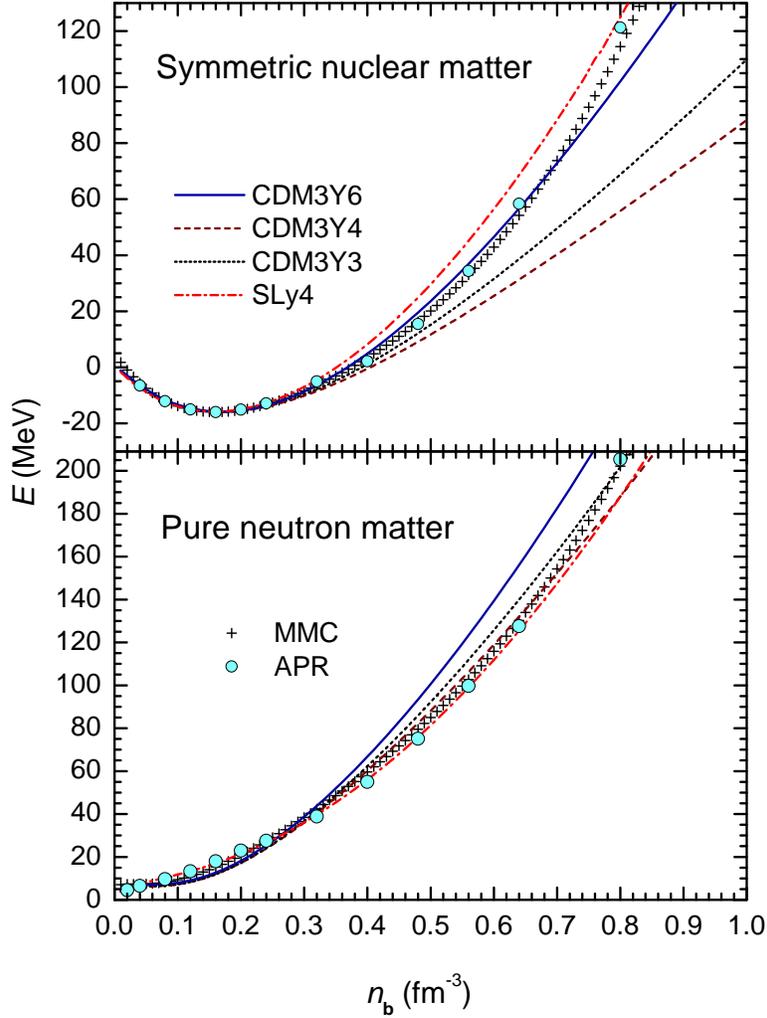}\vspace*{-1cm}
 \caption{(Color online) Energy of the symmetric NM and pure neutron
matter calculated in the HF approximation (\ref{e2}) - (\ref{e3}) using the in-medium
NN interactions that give a \emph{stiff} behavior of $S(n_{\rm b})$ as shown in
upper panel of Fig.~\ref{f1}. The circles and crosses are results of the ab-initio
calculation by Akmal, Pandharipande and Ravenhall (APR) \cite{Ak98} and
microscopic Monte Carlo (MMC) calculation by Gandolfi {\it et al.} \cite{Gan10},
respectively.} \label{f2}
\end{figure}
The main characteristics of the EOS's obtained with these two groups of
density-dependent NN interactions have been discussed in details in
Ref.~\cite{Tha09}. We note here that most of the microscopic calculations
of NM like the ab-initio APR results \cite{Ak98} or the Monte Carlo calculation
by Gandolfi {\it et al.} \cite{Gan10} do predict a stiff behavior of the nuclear
symmetry energy, excepting perhaps the microscopic study by Wiringa {\it et al.}
\cite{Wir88} that predicted a soft behavior of $S(n_{\rm b})$ using the Argonne or
Urbana NN interaction plus a three-nucleon interaction term. The stiff behavior
is also predicted by the recent microscopic Brueckner-Hartree-Fock (BHF) or
Dirac-Brueckner-Hartree-Fock calculations of NM that include the higher-order
many-body correlations and/or three-body forces \cite{Kl06,DBHFa,Li06} as well
as by the latest relativistic mean-field studies \cite{Aru04,Tod05,Pie09}.
\begin{figure}[bht] 
\includegraphics[width=0.7\textwidth]{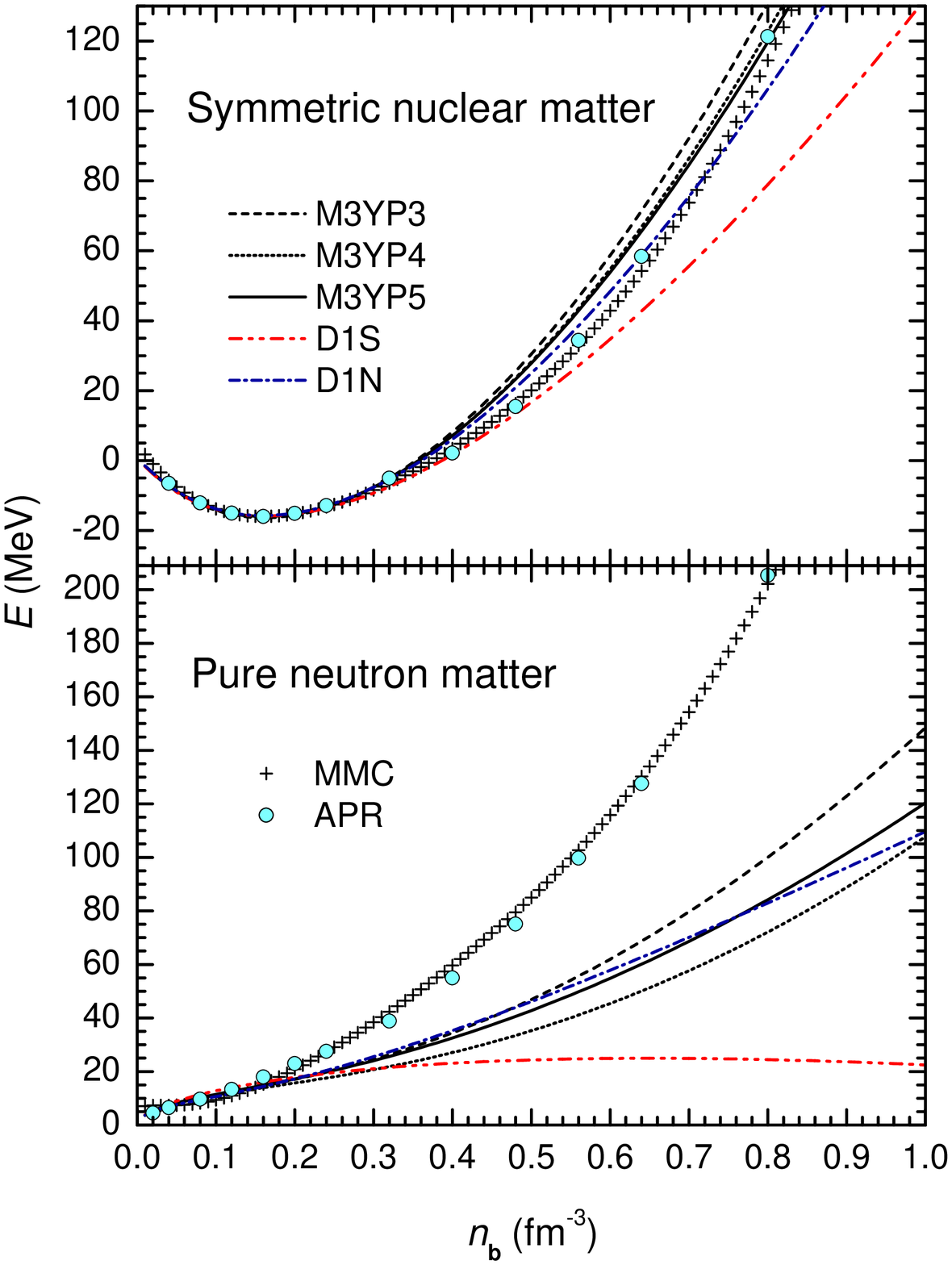}\vspace*{-1cm}
 \caption{(Color online) The same as Fig.~\ref{f2} but obtained with the in-medium
NN interactions that give a \emph{soft} behavior of $S(n_{\rm b})$ as shown in
lower panel of Fig.~\ref{f1}.} \label{f3}
\end{figure}
Because the isovector density dependence of the CDM3Y$n$ interactions has been
parametrized \cite{Kho07} to reproduce simultaneously the BHF results for the isospin-
and density dependent nucleon optical potential by the JLM group \cite{Je77} and
the charge exchange \pn data for the isobaric analog excitation  \cite{Kho07}, the
\emph{stiff} behavior of $S(n_{\rm b})$ given by the CDM3Y$n$ interactions is 
quite close to that given by the BHF calculation. If we simply assume 
the density dependence of the isovector  part of CDM3Y$n$ interactions to be 
the same as that of the isoscalar part, then $S(n_{\rm b})$ has a \emph{soft} 
behavior that has been discussed in our earlier  HF study \cite{Kho96}. 
On the other hand, the isovector density dependence
of the M3Y-P$n$, D1S and D1N interactions were carefully fine tuned against the
structure data observed for a wide range of the neutron (and proton-) dripline
nuclei and the low-density tail of $S(n_{\rm b})$ predicted by these interactions
should be quite realistic. However, there is no physics ground to confirm the validity
of the high-density behavior of $S(n_{\rm b})$ predicted by the \emph{soft}
M3Y-P$n$, D1S and D1N interactions. For the total NM energy (\ref{e3}) that is
sometimes referred to as the nuclear EOS, the HF results given by all considered
interactions for the symmetric NM agree well with the microscopic APR or Monte
Carlo predictions. However, the difference in the symmetry energy lead to very
different behaviors of the energy of pure neutron matter given by the stiff and soft
groups of interactions (see lower panels of Figs.~\ref{f2} and \ref{f3}).
In terms of the NM pressure (\ref{e4}) the soft-type interactions have been shown
\cite{Tha09} unable to reproduce the empirical pressure $P(n_{\rm b})$ of the pure
neutron matter deduced from the HI flow data \cite{Da02}. In the present work we
make such a comparison more accurately, based on the mean-field prediction
for the EOS of the NS matter of the np$e\mu$ composition in the $\beta$-equilibrium
and the recent empirical data deduced from astrophysical measurements of neutron stars
by \"{O}zel {\it et al.} \cite{Ozel10} and Steiner {\it et al.} \cite{Ste10}.

\section{EOS of the $\beta$-stable neutron star matter}

Nuclei in the NS crust are described by the Compressible Liquid Drop Model
by Douchin {\it et al.} (see Ref.~\cite{Dou00} and references therein), using the
model parameters determined with the version SLy4 of the Skyrme interaction
\cite{Ch98}. Within the CLDM, electrons inside the inhomogeneous NS crust
are assumed to form a relativistic Fermi gas. The structure of the NS crust has been
given by the CLDM for the baryon densities up to the \emph{edge} density
$n_{\rm edge}\approx 0.076$ fm$^{-3}$, where a weak first-order phase transition
between the NS crust and liquid (uniform) core takes place \cite{Dou01}.

At baryon densities $n_{\rm b}>n_{\rm edge}$ the NS core is described as
a homogeneous matter of neutrons, protons, electrons and negative muons
($\mu^-$ appear at $n_{\rm b}$ above the muon threshold density,
where  electron chemical potential $\mu_e>m_\mu c^2\approx 105.6$ MeV).
Such a np$e\mu$ composition of the NS core is a realistic assumption up
to the high densities of $n_{\rm b}\simeq 3n_0$. Although the appearance
of hyperons can be expected at higher densities, Douchin and Haensel have
extrapolated their np$e\mu$ model for the EOS of the NS matter up to the
maximum central density (approach used earlier in the ab-initio study by
Akmal {\it et al.} \cite{Ak98}). Thus, the total energy density $\varepsilon$
of the np$e\mu$ matter (including the rest energy of baryons and leptons) is
determined in the present study as
\begin{equation}
\varepsilon(n_{\rm n},n_{\rm p},n_e,n_\mu)=
\varepsilon_{\rm HF}(n_{\rm n},n_{\rm p})+n_{\rm n}m_{\rm n}c^2
+n_{\rm p}m_{\rm p}c^2+\varepsilon_e(n_e)+\varepsilon_\mu(n_\mu)
\label {e5}, \end{equation}
where $\varepsilon_{\rm HF}(n_{\rm n},n_{\rm p})$ is the Hartree-Fock
energy density of nucleons (\ref{e2}); $\varepsilon_e$ and $\varepsilon_\mu$
are the energy densities of electrons and muons, respectively, which are
evaluated in the relativistic Fermi gas model, neglecting electrostatic interaction
\cite{Dou01,Bom01}. The number densities of leptons, $n_e$ and $n_\mu$, are
determined from the charge neutrality condition ($n_{\rm p}=n_e+n_\mu$) and
the relation for the chemical potentials, implied by the $\beta$-equilibrium of the
(neutrino-free) NS matter
\begin{equation}
\mu_{\rm n}= \mu_{\rm p}+\mu_e\  {\rm and}\ \mu_\mu=\mu_e,
\ {\rm where}\  \mu_j=\frac{\partial \varepsilon}{\partial n_j}, \
j={\rm n}, {\rm p}, e, \mu. \label {e6}
\end{equation}
As a result, we can determine uniquely all fractions of the constituent particles
$x_j=n_j/n_{\rm b}$ at the given baryon density $n_{\rm b}=n_{\rm n}+n_{\rm p}$.
Below the muon threshold density ($\mu_e<m_\mu c^2\approx 105.6$ MeV)
the charge neutrality condition leads to the following relation \cite{Bom01}
\begin{equation}
 3\pi^2(\hbar c)^3n_{\rm b}x_{\rm p}-\hat\mu^3=0,\ {\rm where}\
 \hat\mu=\mu_{\rm n}-\mu_{\rm p}=2\frac{\partial E}{\partial\delta}
 \Bigr |_{n_{\rm b}}. \label {e7}
 \end{equation}
Using the total NM energy $E$ given by the HF calculation (\ref{e3}), the density
dependence of the proton fraction at the $\beta$-equilibrium,
$x_{\rm p}(n_{\rm b})$, is readily obtained from the solution of Eq.~(\ref{e7}).
If we assume the parabolic approximation and neglect the contribution from
higher-order terms in (\ref{e3}), then $x_{\rm p}(n_{\rm b})$ is given by the
solution of the well-known equation \cite{Lat04}
\begin{equation}
 3\pi^2(\hbar c)^3n_{\rm b}x_{\rm p}-[4S(n_{\rm b})(1-2x_{\rm p})]^3=0, \label {e8}
 \end{equation}
which shows the crucial role of the NM symmetry energy in the determination of the
proton abundance in the NS matter.

As the baryon density exceeds the muon threshold density, where $\mu_e>
m_\mu c^2\approx 105.6$ MeV, it is energetically favorable for electrons
to convert to negative muons and the charge neutrality condition leads now to the
relation \cite{Bom01}
\begin{equation}
 3\pi^2(\hbar c)^3n_{\rm b}x_{\rm p}-\hat\mu^3-
 [\hat\mu^2-(m_\mu c^2)^2]^{3/2}\theta(\hat\mu-m_\mu c^2)=0, \label {e9}
 \end{equation}
where $\theta(x)$ is the Heaviside step function. Based on the solutions of
Eqs.~(\ref{e7}) and (\ref{e9}), the EOS of the np$e\mu$ matter is fully determined by
the mass density $\rho(n_{\rm b})$ and total pressure $P(n_{\rm b})$ inside
the neutron star
\begin{equation}
\rho(n_{\rm b})=\varepsilon(n_{\rm b})/c^2,\ P(n_{\rm b})=n_{\rm b}^2
{\frac{\partial}{{\partial n_{\rm b}}}}\left[\frac{\varepsilon_{\rm HF}(n_{\rm b})}
 {n_{\rm b}}\right]+P_e+P_\mu. \label {e10}
 \end{equation}

\begin{figure}[bht] \vspace*{3cm}\hspace*{-1cm}
\includegraphics[width=1.1\textwidth]{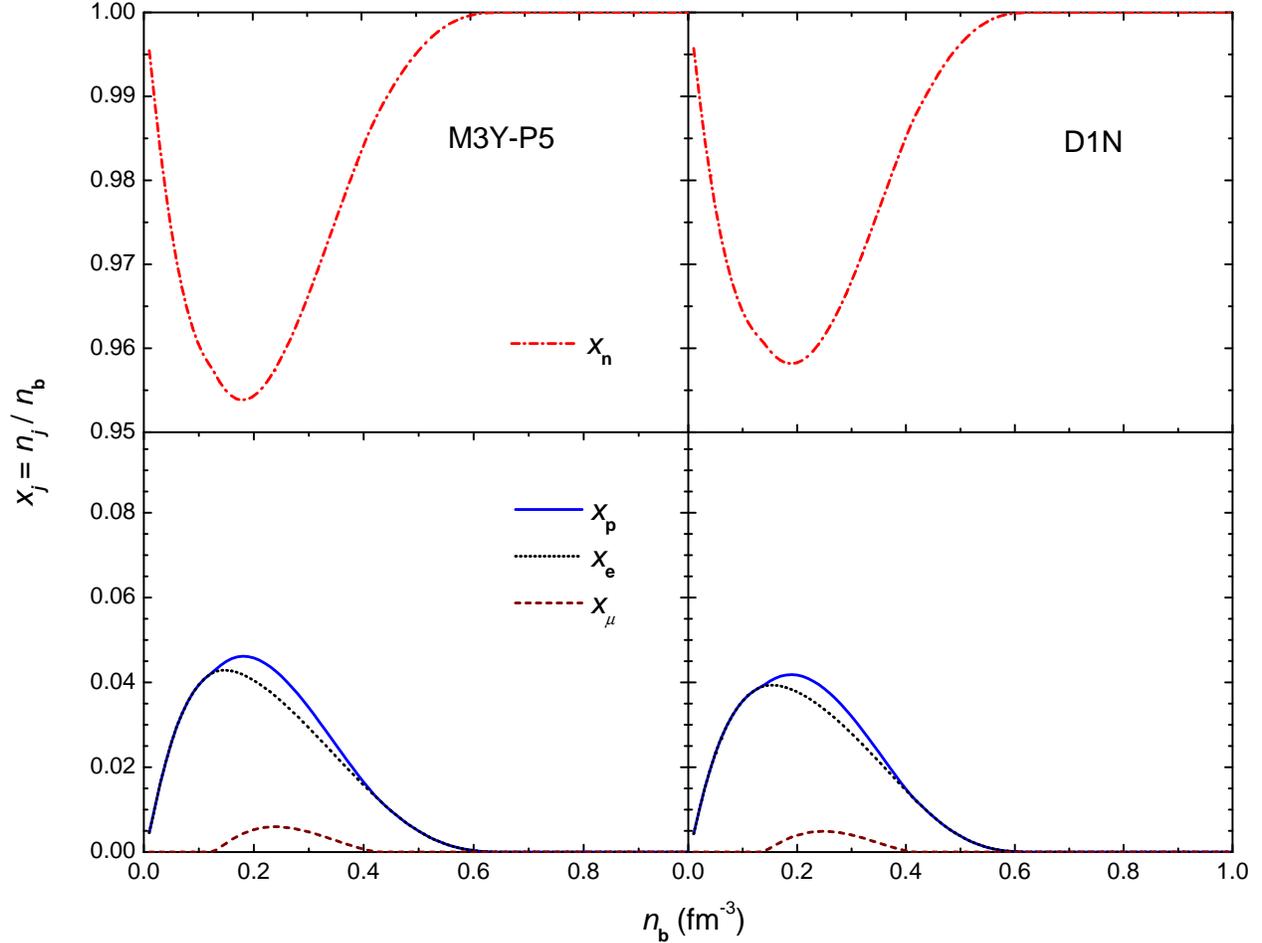}\vspace*{-4cm}
 \caption{(Color online) The fractions $x_j=n_j/n_{\rm b}$ of constituent particles
of the NS matter obtained from the solutions of Eqs.~(\ref{e7}) and (\ref{e9}) using
the mean-field potentials given by the M3Y-P5 and D1N interactions.} \label{f4}
\end{figure}
In the present study, we have first solved Eqs.~(\ref{e7}) and (\ref{e9}) to determine
all fractions $x_j=n_j/n_{\rm b}$ and the EOS of the np$e\mu$ matter
using the total baryon energy $E(n_{\rm b})$ given by the Sly4 interaction.
The accuracy of the numerical procedure was double checked against the published
results for $x_j(n_{\rm b})$ and $P(n_{\rm b})$, obtained with the Sly4
interaction by Douchin and Haensel (tabulated in Refs.~\cite{Dou01,Ioffe}).
Different EOS's of the NS core have been then calculated for the np$e\mu$
matter, using the HF mean-field energies given by different in-medium NN
interactions considered in our study.

\begin{figure}[bht] \vspace*{3.5cm}\hspace*{-1cm}
\includegraphics[width=1.1\textwidth]{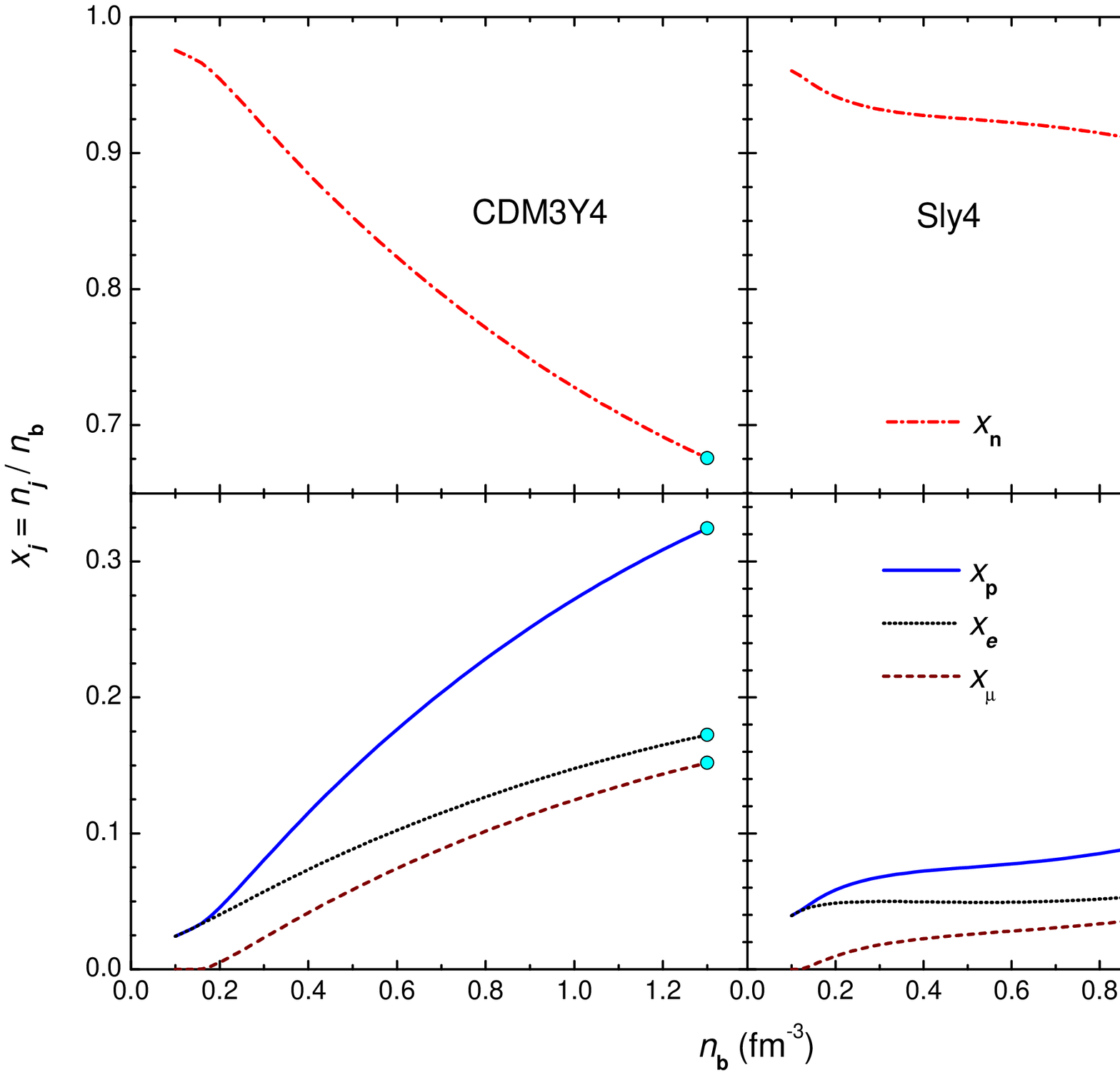}\vspace*{-4cm}
 \caption{(Color online) The same as Fig.~\ref{f4} but using the mean-field
potentials given by the CDM3Y4 and Sly4 interactions. The circles are $n_j$
values calculated at the maximum central densities
$n_{\rm c}$ given by the solution of the TOV equations.} \label{f5}
\end{figure}
Our results for the fractions $x_j=n_j/n_{\rm b}$ obtained with different interactions
are plotted in Figs.~\ref{f4} and \ref{f5}. For the typical soft-type interactions (like
the results obtained for the M3Y-P5 and D1N interactions shown here), the proton and
lepton  fractions are quite small and reach their maxima of around 4\% at  $n_{\rm b}
\approx 0.2$ fm$^{-3}$ (see lower panels of Fig.~\ref{f4}), at exactly the same baryon
density where the symmetry energy $S(n_{\rm b})$ goes through its maximum value.
The fast decrease of $S(n_{\rm b})$ to zero at $n_{\rm b}\approx 0.6-0.7$ fm$^{-3}$
leads also to a drastic decrease of the proton and lepton components in the NS matter
that then becomes \emph{$\beta$-unstable}, pure neutron matter at
$n_{\rm b}> 0.6$ fm$^{-3}$ (see upper panels of Fig.~\ref{f4}).

The fractions $x_j=n_j/n_{\rm b}$ obtained with the stiff-type interactions (see
the results obtained for the CDM3Y4 and Sly4 interactions shown in Fig.~\ref{f5})
are substantially different from those given by the soft-type interactions. Namely,
the proton and lepton  fractions increase steadily with the baryon density like
the corresponding symmetry energy $S(n_{\rm b})$. For the CDM3Y$n$
interactions, the proton fraction $x_{\rm p}$ is above 30\% at the maximum central
density $n_{\rm c}\approx 1.3$ fm$^{-3}$, and the matter at the NS center
becomes less neutron rich (with $x_{\rm n}< 70\%$). In this case, the
$\beta$-equilibrium  of the charge neutral NS matter is kept throughout the
NS core, with the lepton fractions reaching more than 30\%
at $n_{\rm c}$ (see left-lower panel of Fig.~\ref{f5}). It is interesting to note that
about the same behavior of $x_{\rm p}$ is also predicted by the recent complete EOS
of nuclear matter by Shen {\it et al.} \cite{She11}, constructed for use in astrophysical
simulations.  For the Sly4 interaction, with a less stiff increase of $S(n_{\rm b})$
at large densities (see upper panel of Fig.~\ref{f1}), the maximum proton fraction at
$n_{\rm c}\approx 1.2$ fm$^{-3}$ is only about 12\% and the NS matter is, therefore,
more neutron rich compared to the case of CDM3Y$n$ interactions.
Nevertheless, in the case of Sly4 interaction the NS matter remains always
in the $\beta$-equilibrium  \cite{Dou01}.

\begin{figure}[bht] \vspace*{2.5cm}\hspace*{-1cm}
\includegraphics[width=0.88\textwidth]{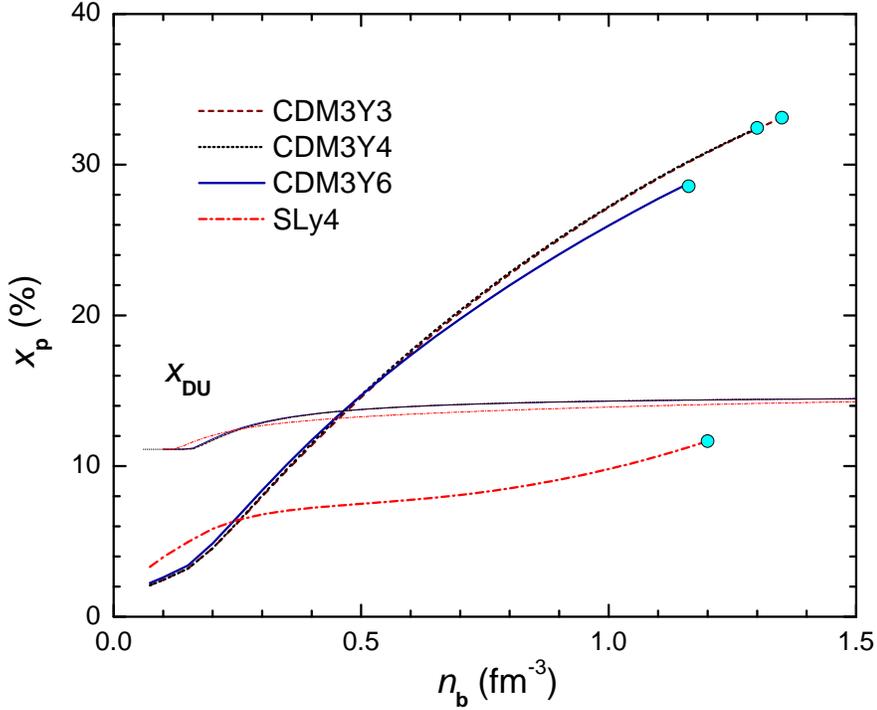}\vspace*{-4cm}
 \caption{(Color online) The proton fraction $x_{\rm p}$ of the $\beta$-stable NS matter
obtained from the solutions of Eqs.~(\ref{e7}) and (\ref{e9}) using the mean-field
potentials given by the stiff-type CDM3Y$n$ and Sly4 interactions. The circles
are $n_{\rm p}$ values calculated at the maximum central densities $n_{\rm c}$
given by the TOV equations. The thin lines are the corresponding DU
thresholds (\ref{e11}). } \label{f6}
\end{figure}
As already discussed in Ref.~\cite{Tha09}, the behavior of the density
dependence of the proton fraction $x_{\rm p}(n_{\rm b})$ plays a very
important role in the determination of the NS cooling rate. In particular,
the powerful direct Urca (DU) process of neutrino emission is allowed only
if the Fermi momenta of the constituent particles in the np$e\mu$ matter
satisfy the triangle conditions \cite{Lat91} that lead to the existence of
a DU threshold $x_{\rm DU}$ for the proton fraction that can be estimated
\cite{Kl06} as
\begin{equation}
x_{\rm DU}=\frac{1}{1+\left(1+r_e^{1/3}\right)^3}, \label {e11}
 \end{equation}
where $r_e=n_e/(n_e+n_\mu)$ is the leptonic electron fraction. $x_{\rm DU}$
has its lowest value of 11.1\% at $r_e=1$ that corresponds to the muon-free
threshold for DU process \cite{Lat94}. It can be concluded immediately from
lower panel Fig.~\ref{f4} that the DU process is \emph{not} possible for the
NS matter generated with the soft-type interactions when $x_{\rm p}$ can reach
at most 4\% and then decreases quickly to zero at $n_{\rm b}> 0.6$ fm$^{-3}$.
Such a (mean-field) challenge to the soft-type in-medium NN interactions has been
pointed out and discussed in Ref.~\cite{Tha09}. Fig.~\ref{f6} shows that for the
stiff-type CDM3Y$n$ interactions, the proton fraction becomes larger than the DU
threshold at a rather modest threshold density of $n_{\rm b}\approx 0.45$ fm$^{-3}$
that is far below the corresponding maximum central densities of $1.2-1.3$ fm$^{-3}$.
Therefore, the DU process should be a realistic scenario for the NS cooling if the
mean-field potential is generated with the CDM3Y$n$ interactions.
For the Sly4 interaction, the $x_{\rm p}$ value remains well below the DU threshold
up to the maximum central density determined by the TOV equations and the DU process
is thus not likely for the NS matter generated with the Sly4 interaction \cite{Dou01}.

\begin{figure}
\includegraphics[width=0.7\textwidth]{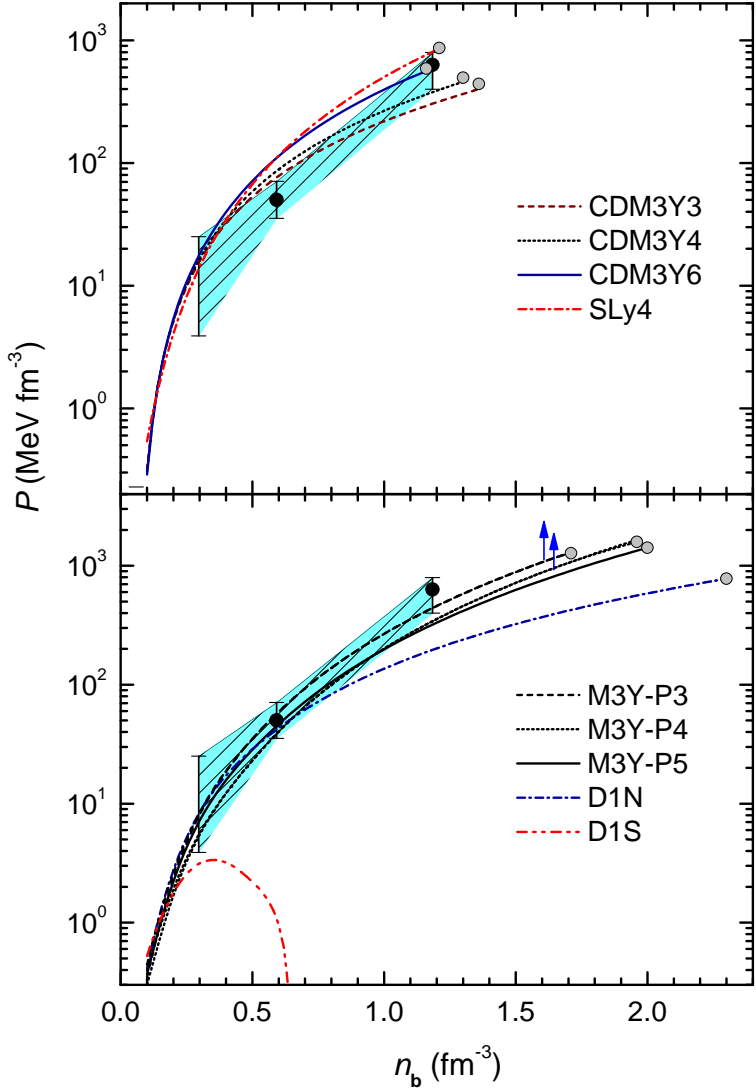}\vspace*{-1cm}
 \caption{(Color online) The pressure inside the NS matter obtained with the in-medium
NN interactions that give \emph{stiff} (upper panel) and \emph{soft} (lower panel)
behavior of $S(n_{\rm b})$, in comparison with the empirical data points
deduced from the astronomical observation of neutron stars \cite{Ozel10}.
The shaded band shows the uncertainties associated with the data determination.
The circles are $P$ values calculated at the corresponding maximum central
densities given by the TOV equations, and the vertical arrows indicate the baryon
densities above which the NS matter predicted by the M3Y-P3 and M3Y-P4
interactions becomes superluminal (see Fig.~\ref{f10} below).} \label{f7}
\end{figure}
We discuss now our results obtained for the total pressure $P(n_{\rm b})$ inside
the neutron star determined by relation (\ref{e10}), which defines the EOS of the
NS matter to be used in the TOV equations. The NS pressure obtained with the
HF mean-field energies $E(n_{\rm b})$ given by different in-medium
NN interactions are plotted in Fig.~\ref{f7}. Since the lepton pressure
($P_e+P_\mu$) inside the NS is about one order of magnitude weaker than
the baryon pressure $P_{\rm b}$, the results shown is Fig.~\ref{f7} are determined
predominantly by $P_{\rm b}$.  One of the main constraints for the NS matter
is that the pressure must satisfy relation $dP/dn \gtrsim 0$ to ensure
the NS matter stability \cite{Glen2}. This (microscopic) stability condition also is
known as le Chatelier's principle \cite{Bom01}.
One can see in lower panel of Fig.~\ref{f7} that $P$ obtained with the
D1S version of Gogny interaction does not comply with such a constraint and this
interaction should not be used to generate the EOS of the NS matter at densities
$n_{\rm b} \gtrsim  2 n_0$. This results stresses again that there could be a plethora
of systematic uncertainties in different models of in-medium NN interaction
that are not visible at low nuclear densities, and the success of any interaction
in the nuclear structure study is not sufficient to ensure its extrapolation to
supranuclear densities.

With the advance in both the astrophysical techniques and modeling of the
NS structure, it became recently feasible to empirically deduce the pressure
of the NS matter at supranuclear densities \cite{Ozel10}. The empirical pressure
determined from the masses and radii observed for the binaries 4U 1608-248,
EXO 1745 -248, and 4U 1820-30 are plotted in Fig.~\ref{f7} as the three
data points spanned by a shadow region of the uncertainties associated with the
data determination \cite{Ozel10}. From Fig.~\ref{f7} one can see that both groups
of in-medium NN interaction agree more or less with the data for NS pressure, while
a similar comparison \cite{Tha09} of the pressure calculated for pure neutron
matter with the empirical value deduced from the HI flow data \cite{Da02} seemed
to favor the stiff-type interactions. In order to distinguish more clearly the EOS's
given by these two groups of interaction, we have further used them in the input
of the Tolman-Oppenheimer-Volkov equations to study the main NS properties.

\section{Neutron star properties}
Different sets of predicted mass density $\rho$ and total pressure $P$ inside
the neutron star (\ref{e10}) have been further used to solve the well-known
Tolman-Oppenheimer-Volkov equations
\begin{eqnarray}
\frac{dP}{dr}&=&-G\frac{m\rho}{r^2}\left(1+\frac{P}{\rho c^2}\right)\left(1+
\frac{4\pi Pr^3}{mc^2}\right)\left(1-\frac{2Gm}{rc^2}\right)^{-1}, \nonumber\\
\frac{dm}{dr}&=&4\pi r^2\rho, \label {e12}
 \end{eqnarray}
where $G$ is the universal gravitational constant,  $r$ is the radial coordinate in the
Schwarzschild metric, and $m$ is the gravitational mass enclosed within the sphere
of radius $r$. The TOV equations (\ref{e12}) are supplemented with the following
equation determining the number of baryons $a$ inside this sphere \cite{Dou01}
\begin{equation}
\frac{da}{dr}=4\pi r^2n_{\rm b}\left(1-\frac{2Gm}{rc^2}\right)^{-1/2}.
\label {e13} \end{equation}
Eqs.~(\ref{e12}), (\ref{e13}) have been integrated from the NS center, with
the boundary conditions at $r = 0:\ P(0) = P_{\rm c},\ m(0) = 0,\ \rho(0)=
\rho_{\rm c}$, and $a(0) = 0$. The stellar surface at $r = R$ is determined
from the boundary condition $P(R) = 0$. The total gravitational mass and
total number of baryons are then determined as $M= m(R),\ A = a(R)$,
respectively. As a result, with different inputs for the NM pressure, the
corresponding solutions of the TOV equations give different NS models in terms
of \emph{one-parameter} families \cite{Dou01} that can be labeled by the
central pressure $P_{\rm c}$  or equivalently by the central density
$\rho_{\rm c}$ of the neutron star.

\begin{figure}[bht] 
\includegraphics[width=0.7\textwidth]{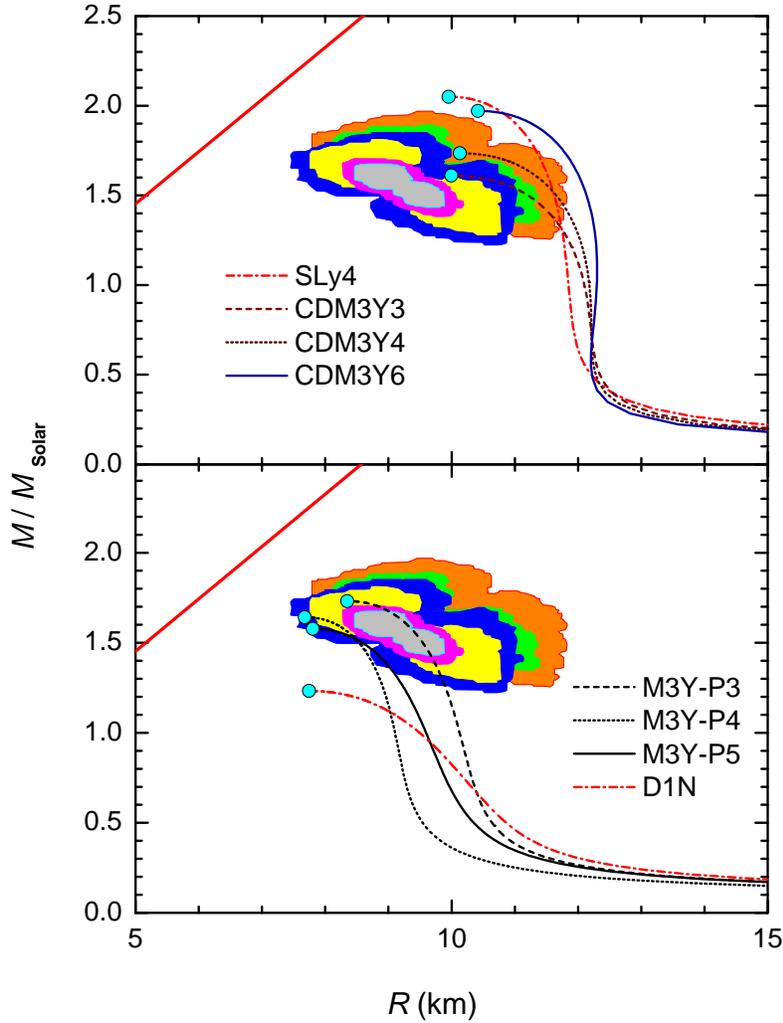}\vspace*{-1cm}
 \caption{(Color online) The NS gravitational mass versus its radius obtained with the
EOS's given by the stiff-type (upper panel) and soft-type (lower panel) in-medium NN
interactions, in comparison with the empirical data (shaded contours) deduced by  \"Ozel
{\it et al} \cite{Ozel10} from  recent astronomical observations of neutron stars.
The circles are values calculated at the maximum central densities. The thick solid (red)
line is the limit allowed by the General Relativity \cite{Glen2}.} \label{f8}
\end{figure}
Thus, at each central density we can uniquely determine the corresponding
gravitational mass $M$ and radius $R$, and the behavior of $M$ versus $R$ is
often used to compare with the measured masses and radii of neutron stars. Our results
obtained with different EOS's are plotted in Fig.~\ref{f8}. The recently measured
masses and radii for the binaries 4U 1608-248, EXO 1745 -248, and 4U 1820-30
\cite{Ozel10} are plotted in Fig.~\ref{f8} as the shaded contours. One can see that
all mass-radius curves lie well below the limit allowed by the General Relativity
\cite{Glen2} and go through or very closely nearby the empirical contours
spanned by the data, excepting the curve given by D1N version of the Gogny
interaction that lies well below the data.
\begin{figure}[bht] 
\includegraphics[width=0.7\textwidth]{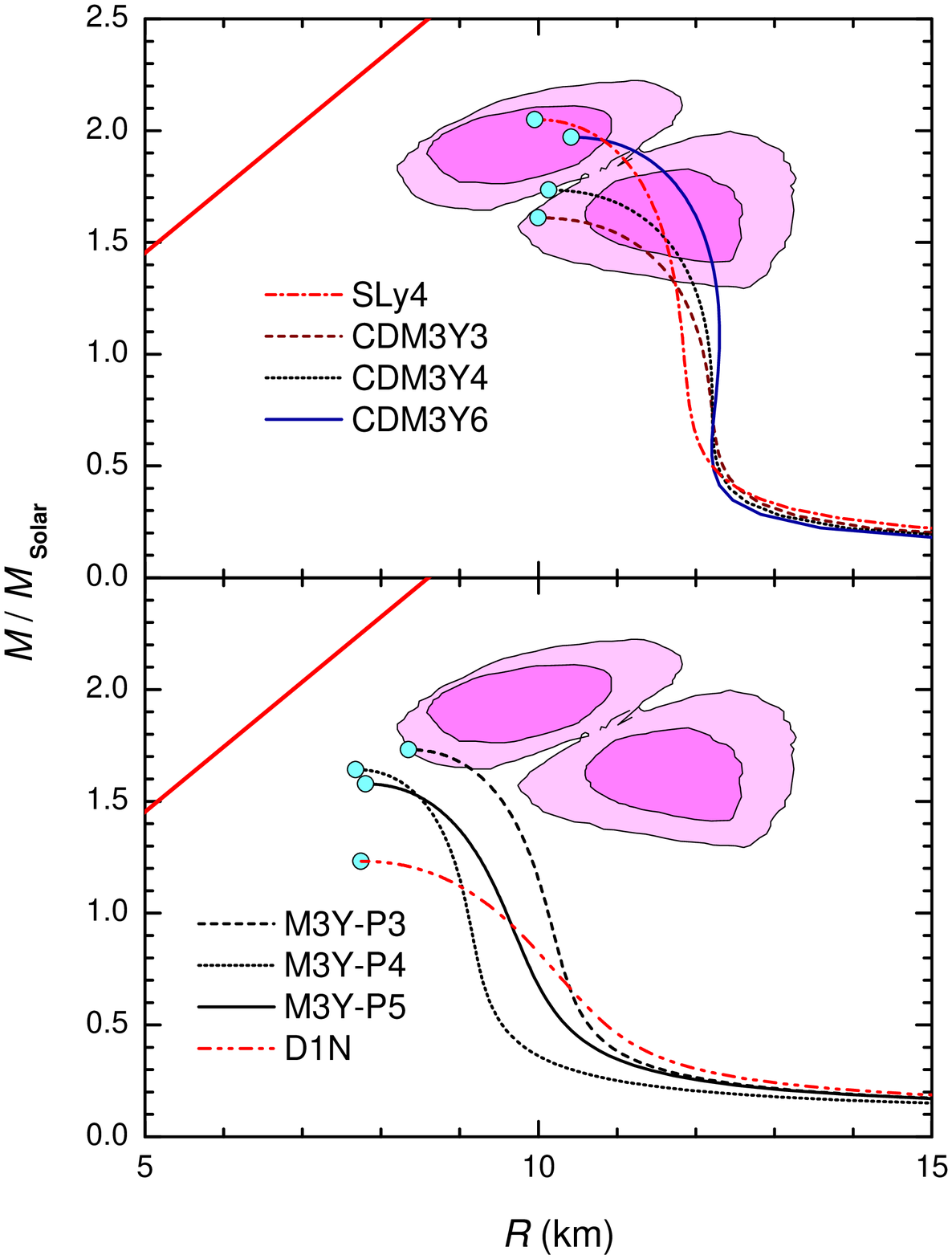}\vspace*{-1cm}
 \caption{(Color online) The same as Fig.~\ref{f8}, but in comparison with the
 empirical data (shaded contours) deduced by Steiner {\it et al.} \cite{Ste10} from
the observation of the X-ray burster 4U 1608-52.} \label{f8a}
\end{figure}
We note that the same EOS for the NS crust (generated in the CLDM
using model parameters determined by the SLy4 interaction \cite{Dou01,Dou00})
has been used in the input of the TOV equations (\ref{e12}). Therefore, the
difference found between the mass-radius curves shown in Fig.~\ref{f8} is
entirely due to the different choices of the in-medium NN interaction used to
generate the EOS of the NS core.  The maximum gravitational masses $M_{\rm G}$
given by the EOS's under study are plotted in Fig.~\ref{f8} as solid circles and they
agree more or less with the recent data. While all the stiff-type interactions give the
corresponding NS radius $R_{\rm G}$ quite close to the empirical range around
10 km, the $M_{\rm G}$ values given by the Sly4 and CDM3Y6 interactions
are slightly higher than the observed masses, close to about twice the solar mass
($M_\odot$). We note, however, that the NS matter in the present study has been
assumed to consist only of baryons, electrons and muons. At high baryon densities
($n_{\rm b}\gtrsim 3n_0$) the hyperons are expected to appear, and the maximum
NS mass becomes then smaller \cite{Lat04,Glen2}. In this case, the nucleon matter
generated with the Sly4 or CDM3Y6 interactions seems well suitable for the
description of the baryon component of the NS matter, while the inclusion of
hyperons might pull the $M_{\rm G}$ values given by the soft-type M3Y-P$n$
interaction to values lying below the empirical boundaries shown in
Fig.~\ref{f8} for the maximum NS mass. Moreover, the $M_{\rm G}$ values
around $2M_\odot$ are still allowed by a broad systematics of the measured
NS masses \cite{Lat04,Ste10}. For the D1N interaction, the $M_{\rm G}$ value is
still within the broad range of the measured NS masses \cite{Lat04} but the radius
$R_{\rm G}$ is rather small, almost 1 km smaller than the minimum limit for $R$
versus $M$: $R\gtrsim 3.6+3.9(M/M_\odot) $ km \cite{Lat04}.

We stress again that the existing data for the NS mass still allow a wide
range for the realistic $M_{\rm G}$ value, from  the lowest value of 1.25
$M_\odot$  \cite{Pod05} up to around  2 $M_\odot$ \cite{Lat04,Ste10},
and the comparison of the predicted results \emph{for both mass and radius} is,
therefore, vital in testing different EOS's of the NS matter. As another example,
we have also compared in Fig.~\ref{f8a} the results predicted by different EOS's
with the mass-radius data deduced by Steiner {\it et al.} \cite{Ste10} from the
observation of the Type-I X-ray burster 4U 1608-52. One can see that results
given by the EOS's obtained with the stiff-type interactions agree nicely with the
empirical data, while those given by the soft-type interactions clearly disagree
with the considered mass-radius data.  Because the TOV equations are deduced
from the Einstein's general relativistic equations for  a \emph{gravitationally}
bound star in the hydrostatic equilibrium, the results shown in Fig.~\ref{f8}
indicate that the EOS's given by both groups of the in-medium NN interactions
\begin{figure}[bht] 
\includegraphics[width=0.7\textwidth]{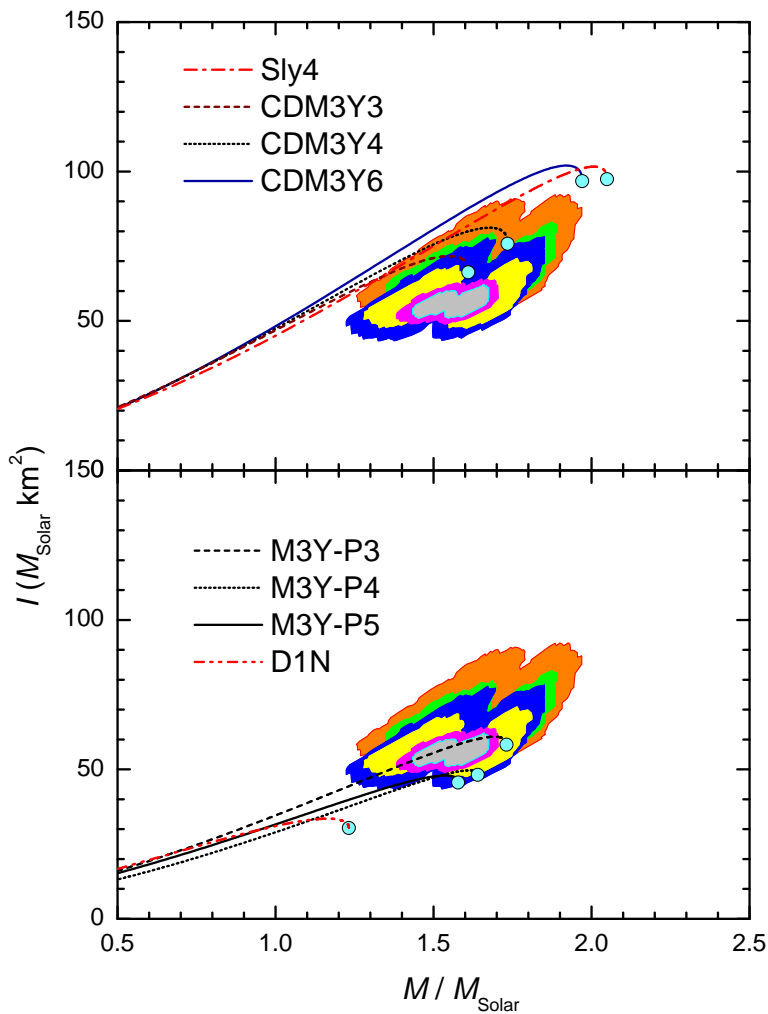}\vspace*{-1cm}
 \caption{(Color online) The NS moment of inertia versus its gravitational mass obtained
with the EOS's given by the stiff-type (upper panel) and soft-type (lower panel)
in-medium NN interactions, in comparison with the empirical data (shaded contours)
deduced from the mass-radius data by  \"Ozel {\it et al} \cite{Ozel10} using
Eq.~(\ref{e14}). The circles are values calculated at the maximum central densities.}
\label{f9}\end{figure}
\begin{figure}[bht] 
\includegraphics[width=0.7\textwidth]{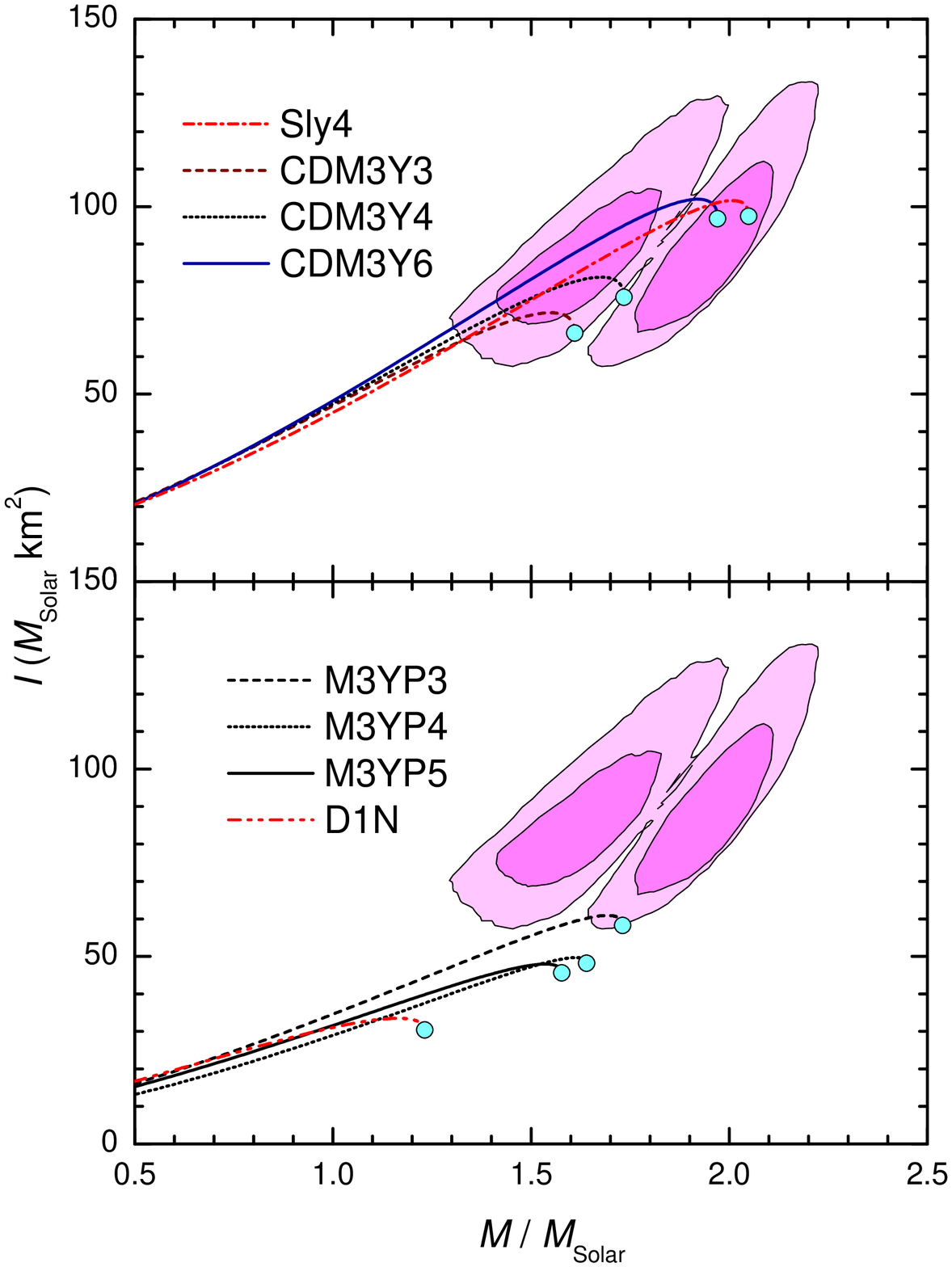}\vspace*{-1cm}
 \caption{(Color online) The same as Fig.~\ref{f9}, but in comparison with the
empirical data (shaded contours) deduced from the mass-radius data by Steiner
{\it et al} \cite{Ste10} for the X-ray burster 4U 1608-52 using Eq.~(\ref{e14}).}
\label{f9a} \end{figure}
(excepting perhaps the D1N interaction) can describe reasonably the empirical
mass-radius data deduced by  \"Ozel {\it et al} \cite{Ozel10}, despite drastically
different proton fractions predicted for the NS matter at supranuclear densities
that lead to very different scenarios for the NS cooling as discussed in the previous
section. Nonetheless, the comparison of our results with the mass-radius data
deduced by Steiner {\it et al.} \cite{Ste10} seems to prefer the stiff-type interactions.

The moment of inertia $I$ of slowly rotating neutron star has been shown \cite{Lat05} as
a good constraint for the EOS of neutron stars and the physics of their interiors.
In general, $I$ can be determined from the Einstein field equations for a compact star
\cite{Wor08}. It has been shown \cite{Lat05,Wor08} that for slowly rotating neutron
stars the expression for $I$ determined from the Einstein field equations can be
approximated quite well by the following empirical relation of the NS mass and radius
\begin{equation}
I\approx (0.237\pm 0.008)MR^2\left[1+4.2\frac{M}{M_\odot}\frac{\rm km}{R}
+90\left(\frac{M}{M_\odot}\frac{\rm km}{R}\right)^4\right].
\label {e14} \end{equation}
Assuming the validity of relation (\ref{e14}), we have transformed the recent
mass-radius data by  \"Ozel {\it et al}  \cite{Ozel10} and Steiner {\it et al}
\cite{Ste10} into the empirical boundaries for realistic values of the
moment of inertia. These new empirical ``data" for the NS moment of inertia $I$
are plotted in Figs.~\ref{f9} and \ref{f9a} by similar shaded contours as in
Figs.~\ref{f8} and \ref{f8a} , and compared with the results obtained with
different EOS's using relation (\ref{e14}). In this way, it is natural to see
that the EOS giving the best agreement with the mass-radius data also give the
corresponding $I$ curves agreeing well with the empirical data for the moment
of inertia. We conclude from the results shown in Figs.~\ref{f8}-\ref{f9a} that the
stiff-type NN interactions give consistently good description to both sets of the
empirical data for the NS masses, radii and moments of inertia.

The main characteristics of the NS configuration determined from the TOV
equations (\ref{e12}), using different EOS's, are given in Table~\ref{t1}. As noted
above, the EOS determined by D1S version of the Gogny interaction gives
\emph{negative} pressure at high baryon densities and violates, therefore, the main
constraint for  a gravitationally bound star \cite{Glen2} in the hydrostatic
equilibrium. As a result, the EOS given by the D1S interaction cannot be used
in the TOV equation (\ref{e12}). From Table~\ref{t1} one can see that the stiff-type
CDM3Y$n$ and Sly4 interactions give the maximum gravitational mass
$1.6\ M_\odot\lesssim M_{\rm G} \lesssim 2 M_\odot$ and radius $R_{\rm G}
\approx 10$ km that are well within the established empirical boundaries as
shown in Figs.~\ref{f8} and \ref{f8a}. We note that both the ab-initio APR calculation
\cite{Ak98} and microscopic Monte Carlo study \cite{Gan10} have obtained
$M_{\rm G} \gtrsim 2 M_\odot$ and the corresponding moment of inertia
$I_{\rm G}$ is, therefore, also somewhat larger than $I_{\rm G}$ values given
by the stiff-type interactions considered here. The soft-type M3Y-P$n$ and, especially,
D1N interactions give the $M_{\rm G}$ and $R_{\rm G}$ values significantly lower
than those given by the stiff-type interactions. Therefore, if hyperons  are included
at high baryon densities, the $M_{\rm G}$ and $R_{\rm G}$ values given by the
soft-type interactions could be well below all the existing empirical estimates.
\begin{table*}
\caption{Configuration of static neutron star given by different NS equations
of state: maximum gravitational mass $M_{\rm G}$, radius $R_{\rm G}$, and
moment of inertia $I_{\rm G}$; maximum central baryon density $n_{\rm c}$,
mass density $\rho_{\rm c}$, and total pressure $P_{\rm c}$; total baryon number $A$;
surface redshift $z_{\rm surf}$; binding energy $E_{\rm bind}$.}
\label{t1}
\begin{tabular}{|c|c|c|c|c|c|c|c|c|c|} \hline
EOS & $M_{\rm G}$ & $R_{\rm G}$ & $n_{\rm c}$ & $\rho_{\rm c}$ & $P_{\rm c}$
   & A   & $ z_{\rm surf}$ & $E_{\rm bind}$ & $I_{\rm G}$ \\
   & $(M_\odot)$ &  (km)   & (fm$^{-3}$) & (10$^{15}$ g/cm$^3$) & (MeV~fm$^{-3}$) &
   (10$^{57}$)  &  & (10$^{59}$ MeV) & ( $M_\odot$~km$^2$) \\ \hline
CDM3Y3 & 1.61 & 10.01 & 1.37 & 2.97 & 444.0  & 2.19  & 0.381 & 2.35 & 66.33  \\
CDM3Y4 & 1.73 & 10.13 & 1.30 & 2.87 & 495.4  & 2.38  & 0.423 & 2.77 & 75.81  \\
CDM3Y6 & 1.97 & 10.42 & 1.17 & 2.66 & 596.0  & 2.76  & 0.506 & 3.65 & 96.78  \\ \hline
M3Y-P3 & 1.73 & 8.36  & 1.71 & 4.10 & 1297.0 & 2.46  & 0.606 & 3.52 & 58.33  \\
M3Y-P4 & 1.64 & 7.72  & 1.96 & 4.72 & 1588.0 & 2.35  & 0.640 & 3.55 & 48.10 \\
M3Y-P5 & 1.58 & 7.81  & 2.00 & 4.78 & 1420.0 & 2.22  & 0.576 & 3.01 & 45.55  \\\hline
D1N    & 1.23 & 7.75  & 2.36 & 5.24 & 819.9 & 1.65  & 0.373 & 1.59 & 30.28  \\\hline
Sly4   & 2.05 & 9.96  & 1.21 & 2.86 & 860.4 & 2.91  & 0.590 & 4.23 & 97.52  \\\hline
CDM3Y3s & 1.13 & 9.36 & 1.61 & 3.26 & 261.1 & 1.47  & 0.246 & 1.12 & 35.62  \\
CDM3Y4s & 1.21 & 9.51 & 1.54 & 3.15 & 275.7 & 1.60  & 0.267 & 1.32 & 40.56  \\
CDM3Y6s & 1.42 & 9.74 & 1.46 & 3.06 & 340.4 & 1.90  & 0.326 & 1.86 & 52.78  \\ \hline
\end{tabular}
\end{table*}

\begin{figure}[bht] \vspace*{2.5cm}\hspace*{-1cm}
\includegraphics[width=0.88\textwidth]{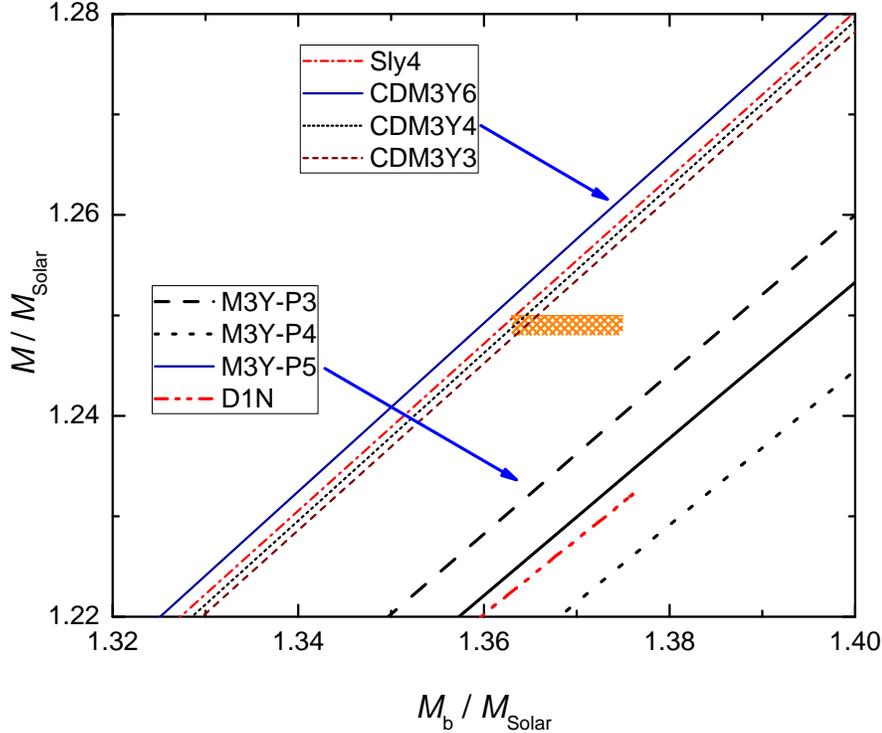}\vspace*{-4cm}
 \caption{(Color online) The gravitational mass $M$ given by different EOS's
of the NS matter plotted versus the corresponding total baryon mass $M_{\rm b}$.
The shaded rectangle is the empirical value inferred from observations
of the double pulsar PSR J0737-3059 by Podsiadlowski {\it et al.} \cite{Pod05}.}
\label{f9b}
\end{figure}
From solutions of the TOV equations at a given radius $R$, the total baryon mass
enclosed within the sphere of radius $R$ can be determined  as $M_{\rm b}(R)= m_N a(R)$,
where $m_N$ is the nucleon mass and the total baryon number $a(R)$ is given by
Eq.~(\ref{e13}). As a result, the $M_{\rm b}(R)$ values are closely correlated with the
corresponding gravitational mass $M(R)$, and difference between these two masses
depends upon the compactness of the neutron star. Furthermore, the low-mass part
of the dependence of gravitational mass on the total baryon mass, $M(M_{\rm b})$,
can be directly compared with the constraint suggested by Podsiadlowski {\it et al.}
\cite{Pod05}. Namely, in a likely scenario that the massive component of the double pulsar
PSR J0737-3059 (the lightest NS observed to date) has been formed by an electron-capture
supernova, the total pre-collapse baryon number of the stellar core (rather precisely
known from the model calculations) has a very small loss of material in the subsequent
collapse. Therefore, the accurate gravitational mass $M=1.249\pm 0.001\ M_\odot$
estimated from the pulsar timing can be used to extract a stringent constraint on the
low-mass part of $M(M_{\rm b})$, which is plotted as the shaded rectangle in
Fig.~\ref{f9b}. As discussed also in Ref.~\cite{Gan10}, a realistic EOS of the NS
matter should give $M(M_{\rm b})$ curve going through or nearby the shaded box
in Fig.~\ref{f9b} in order to be consistent with the observation of the double pulsar
PSR J0737-3059. One can see that the stiff-type NN interactions give $M(M_{\rm b})$
curves passing very close to the left corner of the shaded box, while those given by the
soft-type interactions clearly underestimate the empirical data. At $M\approx 1.25
M_\odot$, the baryon density ($n_{\rm b}> 3n_0$) is well above the hyperon threshold
and the gravitational mass should become smaller if the hyperons are included
\cite{Lat04,Glen2}. This would probably push the $M(M_{\rm b})$ curves given by
the stiff-type interactions right into the shaded box in Fig.~\ref{f9b}. In contrast,
there is no physics mechanism that can push the $M(M_{\rm b})$ curves given
by the soft-type interactions higher up, to be close to the empirical shaded region.

Another  important NS observable is the surface redshift of photons emitted from the
NS photosphere that is determined as
\cite{Dou01}
\begin{equation}
z_{\rm surf}=\left(1-\frac{r_{\rm g}}{R}\right)^{-1/2}-1,\ {\rm where}\
r_{\rm g}=\frac{2GM}{c^2}\approx 2.95 \frac{M}{M_\odot}.
\label {e15} \end{equation}
It is obvious from Eq.~(\ref{e15}) that the measurement of $z_{\rm surf}$
is vital for the determination of the mass/radius ratio. At the typical masses
around $M\approx 1.5 M_\odot$, the stiff-type CDM3Y$n$ interactions give
$z_{\rm surf}\approx 0.30-0.35$ while the soft-type M3Y-P$n$ interactions
give $z_{\rm surf}\approx 0.35-0.40$, which agree reasonably with
the empirical value $z_{\rm surf}\approx 0.35$ deduced from the X-ray spectra of
the burster EXO 0748-676 \cite{Cot02}. Although the maximum $z_{\rm surf}$
value given by the D1N interaction is close to that empirical data, the maximum NS
mass is only around $1.2 M_\odot$, and this effect is also showing up in a poor
agreement of the D1N results with the mass-radius data as illustrated in
Figs.~\ref{f8} and \ref{f9}.

For the NS binding energy, we have used the standard definition of $E_{\rm bind}$
as the mass defect with respect to an unbound configuration consisting of the
same baryon number \cite{Dou01}. Namely, the mass defect with respect to a
dispersed configuration of a pressureless cloud of $^{56}$Fe dust, with mass
per nucleon $m_{\rm Fe}$ = mass of $^{56}$Fe atom divided by 56
\begin{equation}
 E_{\rm bind}=(Am_{\rm Fe}-M)c^2.  \label {e16}
\end{equation}
At the typical masses around $1.5 M_\odot$, all considered interactions
give $E_{\rm bind}$ values within the range of $(2 \div 3)\times 10^{59}$ MeV.
Exception again is the D1N interaction that give a weaker binding energy of
$E_{\rm bind}\approx 1.6\times 10^{59}$ MeV with the maximum gravitational
mass $M_{\rm G}\approx 1.2 M_\odot$.

\begin{figure}[bht] 
\includegraphics[width=0.7\textwidth]{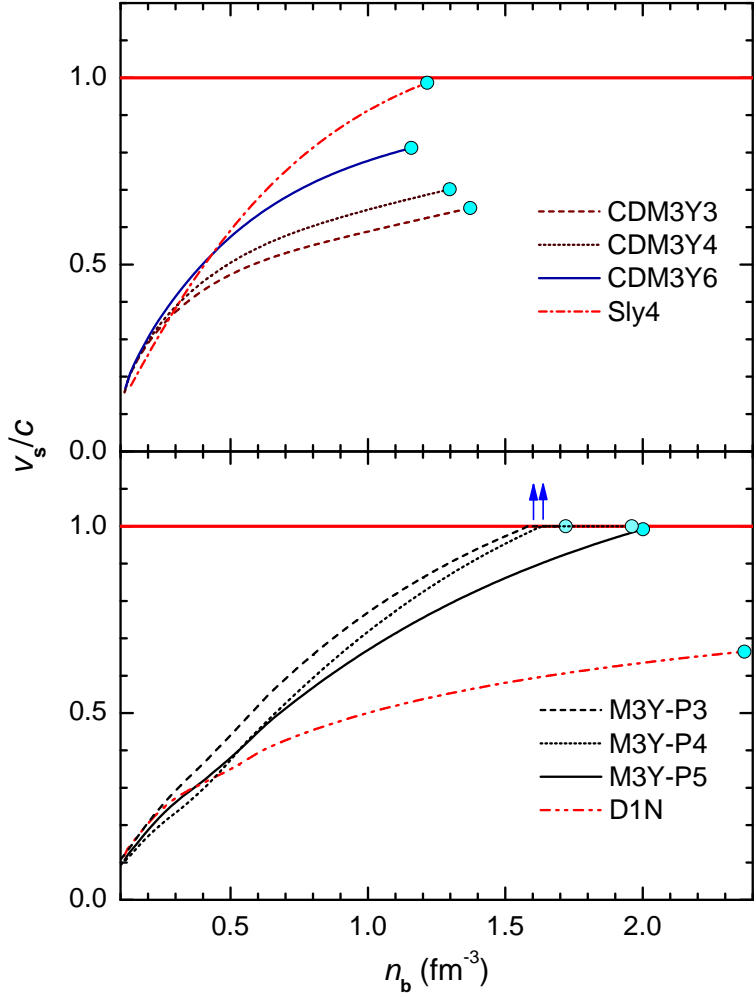}\vspace*{-1cm}
 \caption{(Color online) The adiabatic sound velocity versus baryon density obtained
with the EOS's given by the stiff-type (upper panel) and soft-type (lower panel)
in-medium NN interactions. The thick solid (red) lines are the subluminal limit
($v_{\rm s}\leqslant c$), and the vertical arrows indicate the baryon densities above
which the NS matter predicted by the M3Y-P3 and M3Y-P4 interactions becomes
superluminal (see details in the text).} \label{f10}\end{figure}
Concerning the maximum central pressure, the TOV equations using the EOS's
based on the soft-type M3Y-P$n$ interactions give much too high $P_{\rm c}$
values (see Table~\ref{t1}). The behavior of the central pressure is directly correlated
with the central density. In this context, it is of interest to consider
the \emph{causality} condition \cite{Bom01} that implies the adiabatic sound
velocity in the stellar medium to be \emph{subluminal}, i.e.,
\begin{equation}
 v_{\rm s}=\sqrt{\frac{dP(\rho)}{d\rho}}\leqslant c,  \label {e17}
\end{equation}
where $P(\rho)$ is the total pressure of the NS matter as function of the total mass
density $\rho$. We have estimated the sound velocity $v_{\rm s}$ using the EOS's
given by the two groups of the in-medium NN interactions and the results are plotted
in Fig.~\ref{f10}. While the soft-type M3Y-P$n$ interactions give more or less the
same $v_{\rm s}$ values at low densities as those given by the stiff-type CDM3Y$n$
interactions, EOS given by the M3Y-P3 and M3Y-P4 interactions begins to violate
the causality condition at $n_{\rm b}\approx 1.60$ and 1.64 fm$^{-3}$, respectively.
For further use in the TOV equations, we have assumed in these two cases
a causal EOS \cite{Rho74} given by $P(\rho)=c^2\rho - \epsilon_{\rm C}$ at
the $n_{\rm b}$ larger than 1.6 and 1.64 fm$^{-3}$, respectively, with constant
$\epsilon_{\rm C}$ chosen to ensure the continuity of the energy density across
the critical densities. The M3Y-P5 interaction is doing better and gives
$v_{\rm s}\approx 0.992~ c$ at the maximum central density. For the stiff-type Sly4
interaction, we obtained $v_{\rm s}\approx 0.987~ c$ at the maximum $n_{\rm b}$,
and this value indicate a sound velocity closely comparable to velocity of light in
the dense NS core, as discussed earlier in Ref.~\cite{Dou01}.

In conclusion, we have tested two sets of the in-medium NN interactions in the
description of the main properties of neutron star based on the TOV equations.
We found that all stiff-type interactions are quite realistic in describing
the latest empirical constraints for the EOS of the NS matter, pressure and
mass-radius data. In particular, the CDM3Y6 and Sly4 interactions should be
the appropriate choice for the future NS studies,
when hyperon presence is taken into account at supranuclear densities.
Concerning the soft-type interactions, the EOS's given the M3Y-P3 and M3Y-P4
interactions have been modified at high densities to avoid the violation of
the causality condition. Moreover, the overall agreement of the results
given by the soft-type interactions with the same empirical data is not as good
as that obtained with the stiff-type interactions. It is clear that such effects
are caused not only by a drastic difference in the nuclear symmetry energy alone,
but also by quite different functional structures of the considered
interactions, like, e.g., the \emph{zero-range} density-dependent form of the
M3Y-P$n$ and Gogny interactions versus the \emph{finite-range} form
of the CDM3Y$n$ interactions.

\begin{figure}[bht] 
\includegraphics[width=0.8\textwidth]{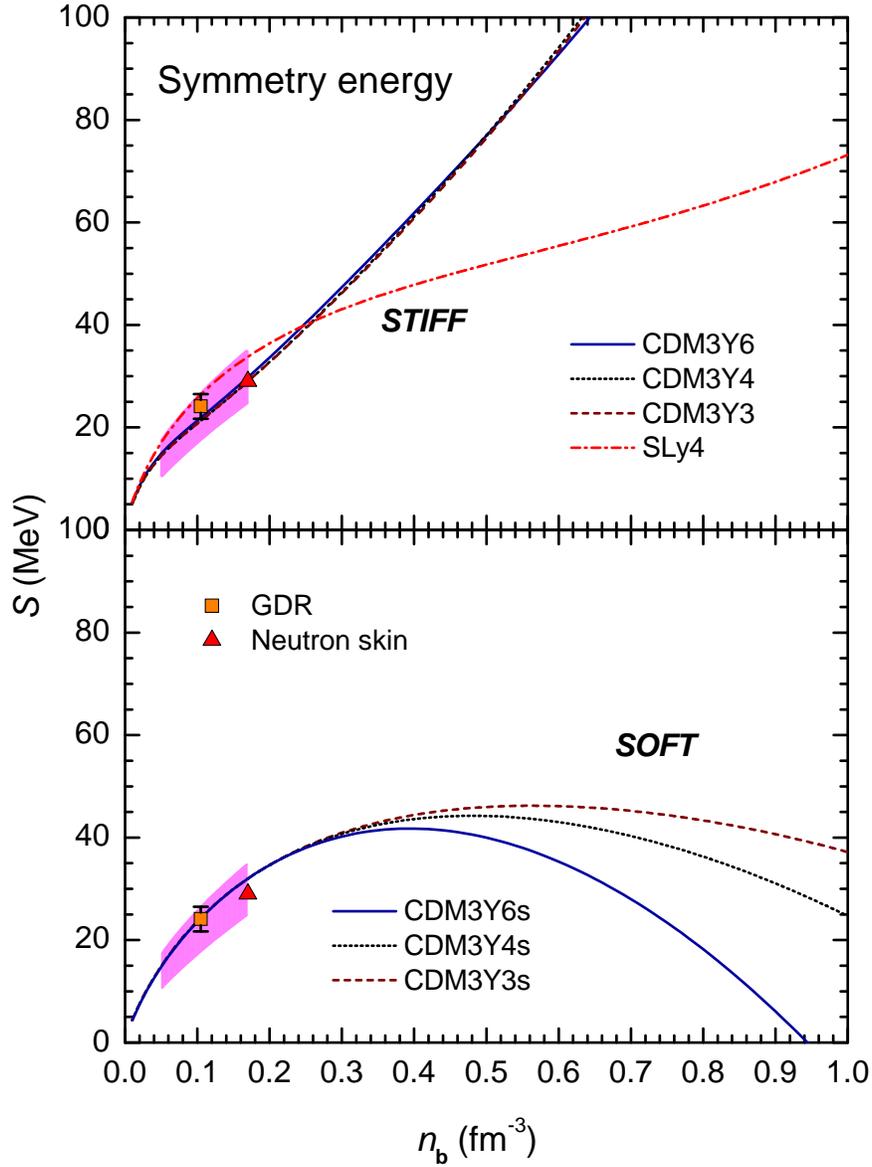}\vspace*{-1cm}
\caption{(Color online) The same as in Fig.~\ref{f1} but $S(n_{\rm b})$ curves in the
lower panel were obtained with the soft CDM3Y$n$s interactions.}
\label{f11} \end{figure}

To explore explicitly the effects caused by the nuclear symmetry energy to the
NS properties in this kind of study, one needs to make a similar analysis with
the HF energy densities (\ref{e2}) obtained essentially with the same in-medium
NN interaction but using different ansatzs for its isospin dependence so that
different behaviors of the symmetry energy $S(n_{\rm b})$ can be tested.
We present here the results of such a test using the CDM3Y$n$ interactions that
have the isoscalar (IS) and isovector (IV) parts determined as
\begin{equation}
 v_{\rm IS(IV)}(n_{\rm b},s)=F_{\rm IS(IV)}(n_{\rm b})v_{\rm IS(IV)}(s),
 \label{g1}
\end{equation}
where $s$ is the internucleon distance. The radial strengths of the IS and IV
interactions $v_{\rm IS(IV)}(s)$ were kept unchanged, as derived from the
M3Y-Paris interaction \cite{An83},
in terms of three Yukawas (see detailed formulas in Ref.~\cite{Kho96}).
The isoscalar density dependence $F_{\rm IS}(n_{\rm b})$ has been parametrized
\cite{Kho97,Kho07r} to reproduce the saturation properties of symmetric NM
in the HF calculation, while the isovector density dependence $F_{\rm IV}(n_{\rm b})$
has been parametrized \cite{Kho07} to reproduce the BHF results for the isospin-
and density dependent nucleon optical potential in the nuclear matter limit \cite{Je77}.
Thus, the stiff behavior of $S(n_{\rm b})$ given by the CDM3Y$n$ interactions is
actually associated with that given by the BHF calculation. In our earlier  HF study
\cite{Kho96} we have simply assumed the IV density dependence to be the same as
that of the IS part, and $S(n_{\rm b})$ has then a typical soft behavior that is illustrated
in Fig.~\ref{f11}. The same assumption has also been used in the NM studies
by Basu {\it et al.} \cite{Basu08} using the density dependent version DDM3Y of the
M3Y-Reid interaction \cite{Be77}.

To study the effects caused by the stiffness of the symmetry energy, we have used
in the present work also a soft version of the CDM3Y$n$ interactions with the IV
density dependence taken as $F_{\rm IV}(n_{\rm b})=1.1F_{\rm IS}(n_{\rm b})$.
The factor of 1.1 has been found \cite{Kho05} to give the best fit of the charge
exchange $^6$He$(p,n)^6$Li data as well as realistic value of the symmetry coefficient
$J=S(n_0)\approx 30$ MeV. These soft CDM3Y$n$ interactions (denoted hereafter
as the CDM3Y$n$s interactions) give a more moderate \emph{soft} behavior of the
symmetry energy compared to the M3Y-P$n$ interactions (compare Figs.~\ref{f1}
and \ref{f11}).
 \begin{figure}[bht] \vspace*{3.5cm}\hspace*{-1cm}
\includegraphics[width=1.1\textwidth]{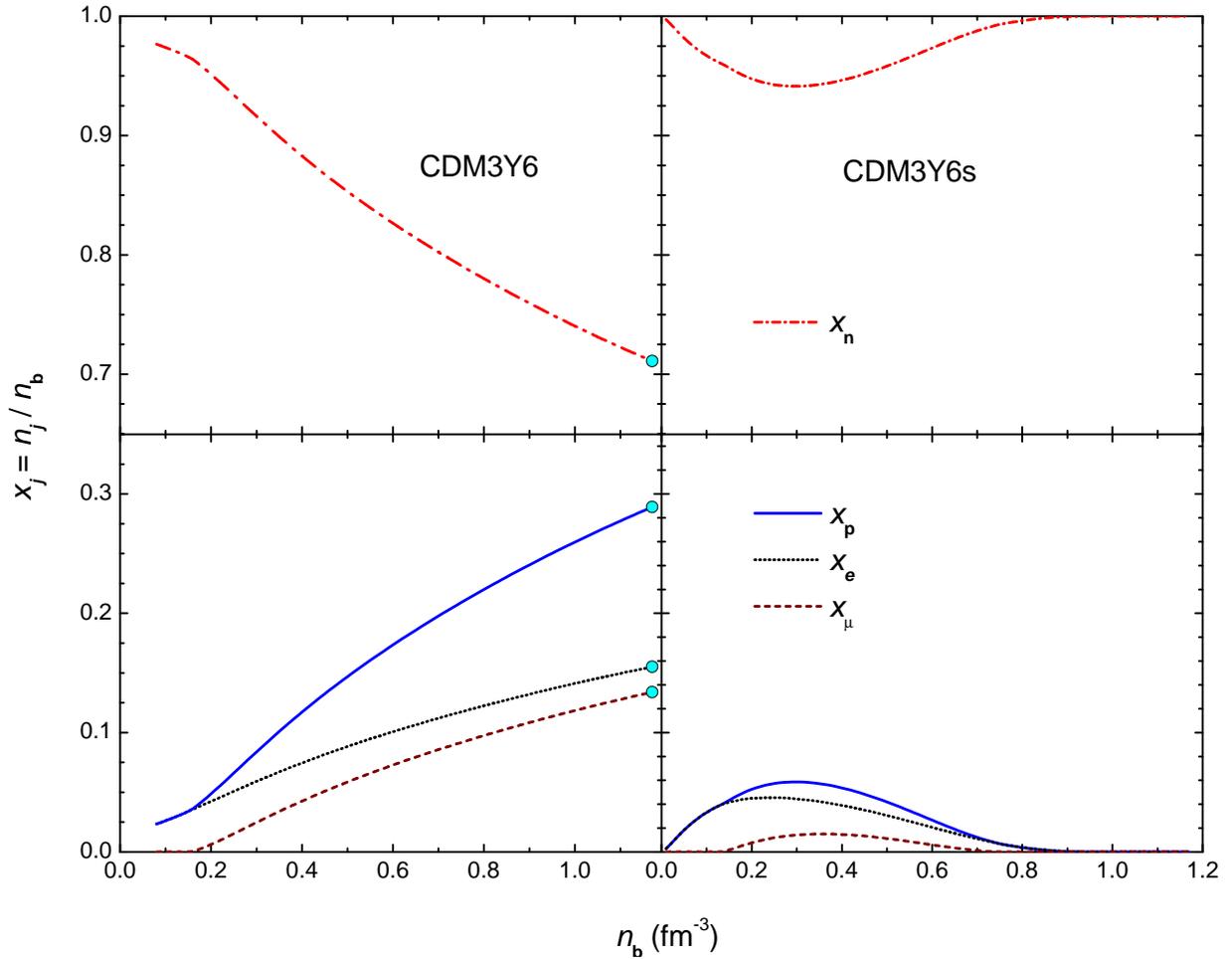}\vspace*{-4cm}
 \caption{(Color online) The same as Fig.~\ref{f4} but using the mean-field
potentials given by the CDM3Y6 and CDM3Y6s interactions.} \label{f12}
\end{figure}
The soft CDM3Y$n$s interactions have been further used to determine the
composition of the $\beta$-stable NS matter by solving Eqs.~(\ref{e7}) and
(\ref{e9}), and the fractions $x_j=n_j/n_{\rm b}$ obtained with the soft
CDM3Y6s interaction are compared with those given by the stiff CDM3Y6
version in Fig.~\ref{f12}. As expected, the proton and lepton  fractions obtained
with the CDM3Y6s interaction are much smaller than those obtained with
the stiff CDM3Y6 interaction. The fast decrease of $S(n_{\rm b})$
to zero at $n_{\rm b}\approx 0.8-0.9$ fm$^{-3}$ leads also to the disappearance
of protons and leptons in the NS matter that soon becomes the $\beta$-unstable,
pure neutron matter at $n_{\rm b}\gtrsim 0.9$ fm$^{-3}$. Like the results obtained
with the soft-type M3Y-P$n$ interactions, the DU scenario of the NS cooling should
also be excluded for the NS matter generated with the soft CDM3Y$n$s interactions,
because $x_{\rm p}$ can reach only around 5 - 6\% and then decreases quickly
to zero at $n_{\rm b}> 0.8$ fm$^{-3}$.

\begin{figure}[bht] 
\includegraphics[width=0.8\textwidth]{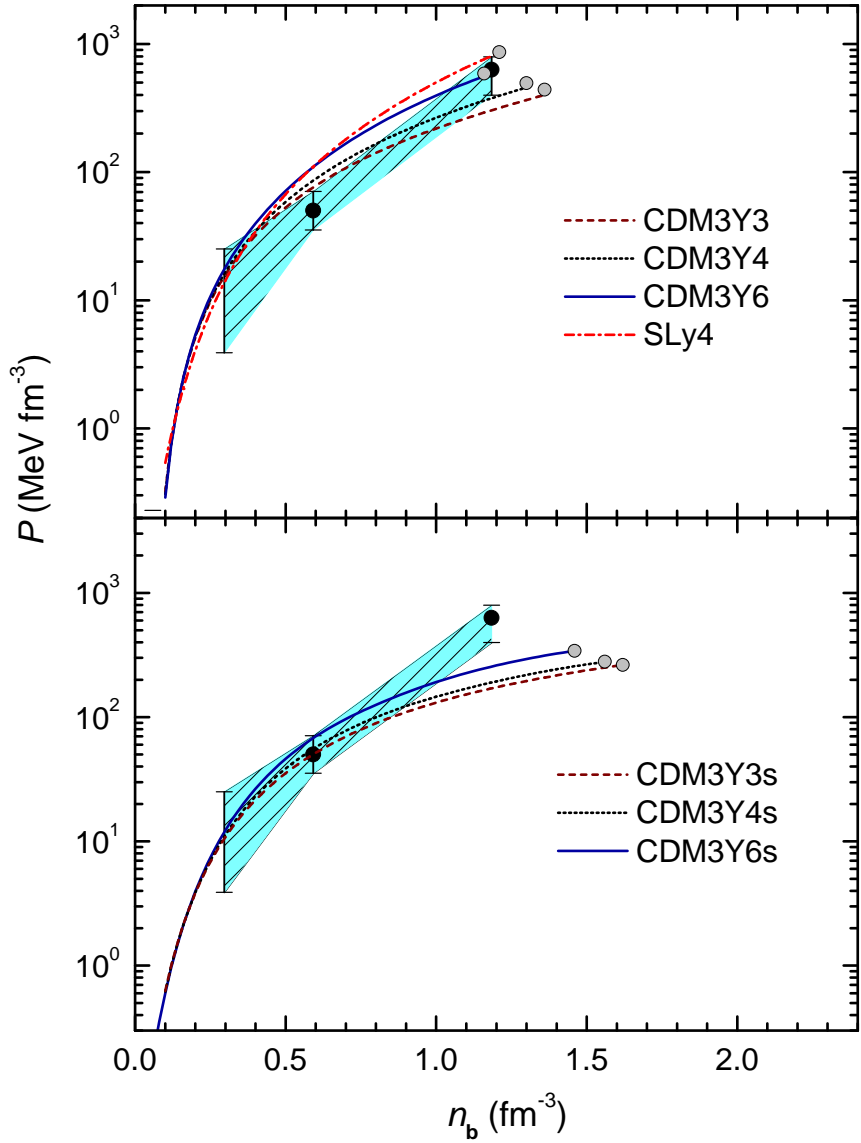}\vspace*{-1cm}
 \caption{(Color online) The same as in Fig.~\ref{f7} but $P(n_{\rm b})$ curves in the
lower panel were obtained with the soft CDM3Y$n$s interactions.} \label{f13}
\end{figure}
The pressures of the NS matter (\ref{e10}) obtained with the two sets of the
CDM3Y$n$ interaction are compared with the empirical data in Fig.~\ref{f13},
and one can see that the soft CDM3Y$n$s interactions fail to account for the
empirical NS pressure at high densities. This effect is well expected because
$P(n_{\rm b})$ is determined from the first derivative of the NM
energy and the decrease of the NM symmetry energy $S(n_{\rm b})$ at
high densities leads to a negative contribution of the symmetry term
of the NM pressure to the total $P(n_{\rm b})$ value.
In other words, the EOS's given by the two sets of the CDM3Y$n$
interaction are substantially different at high baryon densities,
and this effect is entirely due to the different behaviors of the symmetry
energy at high baryon densities.

\begin{figure}[bht] 
\includegraphics[width=0.8\textwidth]{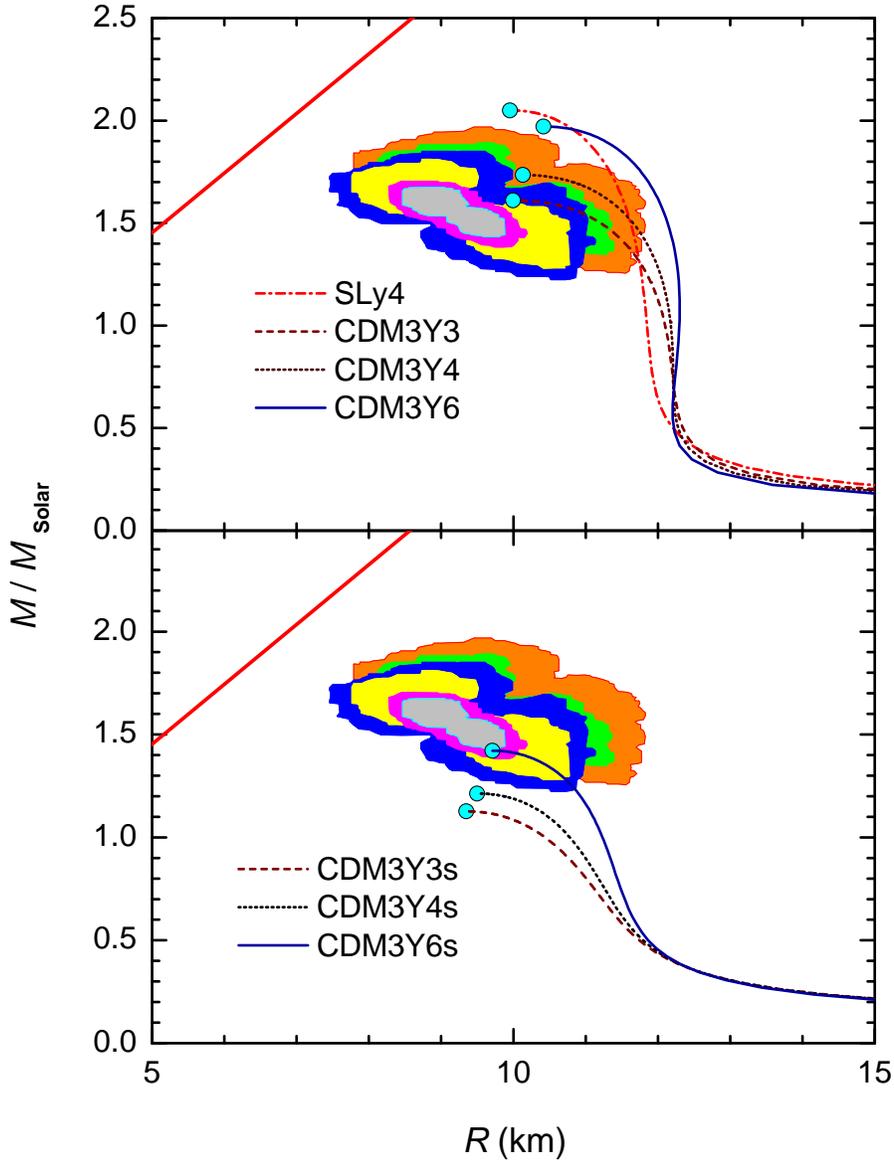}\vspace*{-1cm}
 \caption{(Color online) The same as Fig.~\ref{f8} but mass-radius curves in the
lower panel were obtained with the soft CDM3Y$n$s interactions.} \label{f14}
\end{figure}
\begin{figure}[bht] 
\includegraphics[width=0.8\textwidth]{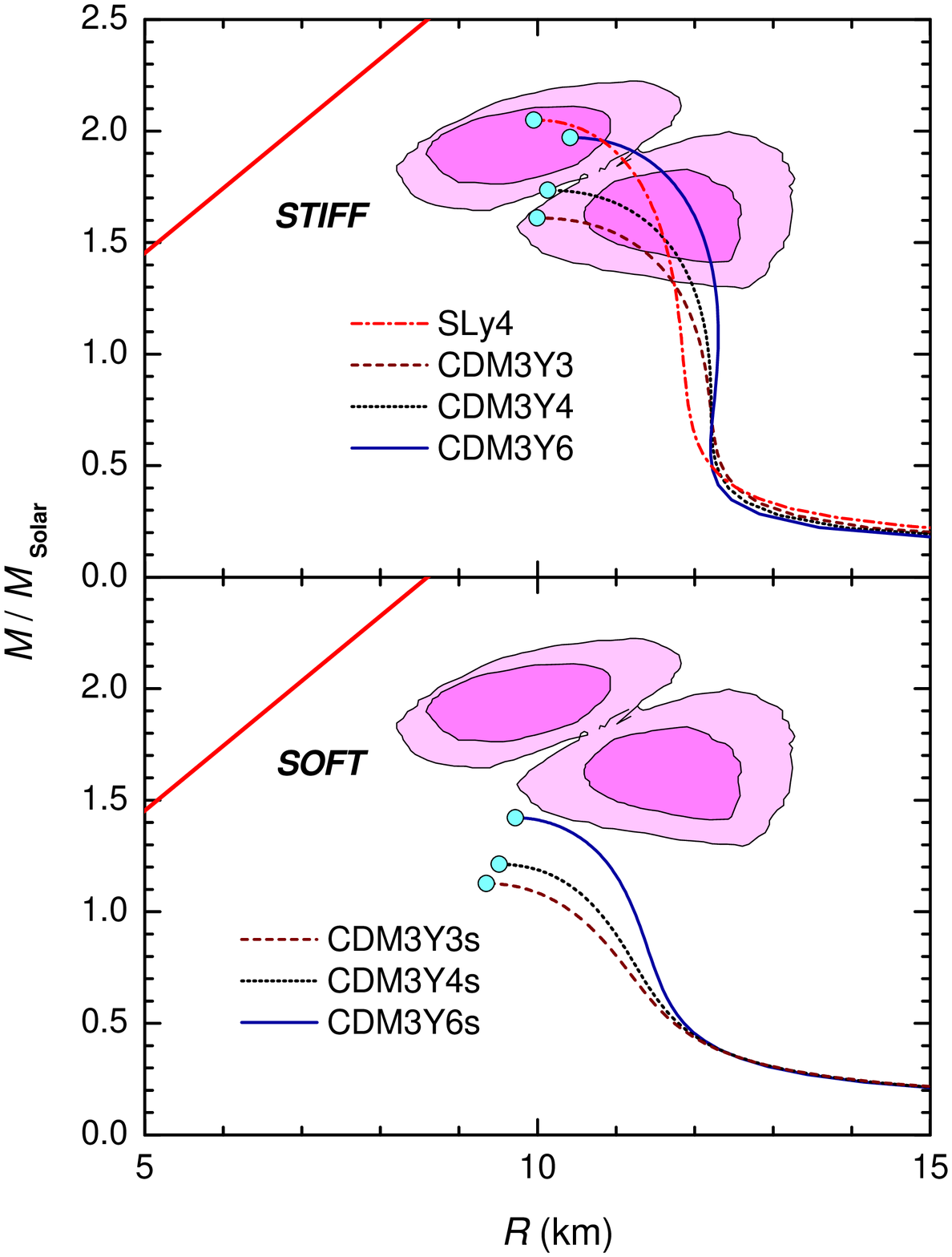}\vspace*{-1cm}
 \caption{(Color online) The same as Fig.~\ref{f8a} but mass-radius curves in the
lower panel were obtained with the soft CDM3Y$n$s interactions.} \label{f15}
\end{figure}
The mass density $\rho$ and total pressure $P$ of the NS matter
(\ref{e10}) obtained with the \emph{soft} CDM3Y$n$s interactions
have been further used as input of the TOV equations (\ref{e12}), and the
obtained NS properties are given in Table~\ref{t1} and illustrated in Figs.~\ref{f14}
and \ref{f15}. The most obvious effect caused by changing slope of the symmetry
energy from \emph{stiff} to \emph{soft} is the reduction of the maximum gravitational
mass $M_{\rm G}$ and radius $R_{\rm G}$ as illustrated in Fig.~\ref{f14}. The
$M_{\rm G}$ value is changing from $1.6\sim 2\ M_\odot$ to a significantly
lower range of $1.1\sim 1.4\ M_\odot$, with a much worse description of the
empirical mass-radius data \cite{Ozel10,Ste10} as shown in Figs.~\ref{f14}
and \ref{f15}. Then, if the hyperons are included at high baryon densities,
the $M_{\rm G}$ values given by the soft CDM3Y$n$s interactions
can be driven to values lying well below the mass of the lightest NS observed
so far ($M=1.25\ M_\odot$) \cite{Pod05}. Together with the maximum central
pressure $P_{\rm c}$, the total baryon number, surface redshift, binding energy
and moment of inertia also become smaller when the EOS of the NS core is obtained
in the HF calculation using the soft CDM3Y$n$s interactions. It is interesting
to note that the EOS's obtained with the soft CDM3Y3s and CDM3Y4s interactions
not only give the NS mass and radius values lying outside the empirical boundaries
but also the surface redshifts $z_{\rm surf}$ clearly in disagreement with the data
($z_{\rm surf}\approx 0.35$) deduced from the X-ray burst spectra of neutron
stars \cite{Cot02}. These results together with those obtained
with the soft-type M3Y-P$n$ and D1N interactions discussed above indicate that
a \emph{soft} density dependence of the NM symmetry energy could cause strong
deviation of the calculated NS properties from the empirical estimates. Although one
still cannot completely rule out the EOS of the NS matter with a \emph{soft}
behavior of the symmetry energy based on this discussion, the results shown in
Figs.~\ref{f13}- \ref{f15} clearly favor the \emph{stiff} behavior and confirm again the
vital role of the nuclear symmetry energy in the theoretical modeling of neutron star.

\section{Summary}
The EOS of the np$e\mu$ matter of neutron star in $\beta$-equilibrium has
been studied in details using the nuclear mean-field potentials obtained in the HF
method with different choices of the effective (in-medium) NN interactions that
give two different behaviors of the NM symmetry energy at supranuclear densities
(the \emph{soft} and \emph{stiff} scenarios).
The fast decrease of the soft NM symmetry energy to zero at
$n_{\rm b}\approx 0.6-0.7$ fm$^{-3}$ results on a drastic decrease of the
proton and lepton components in the uniform NS core that then becomes the
\emph{$\beta$-unstable}, pure neutron matter at $n_{\rm b}> 0.6$ fm$^{-3}$.
Very small proton fraction in the NS matter given by the soft-type interactions
excludes the direct Urca process in the NS cooling, whereas the DU process is
well possible for the \emph{$\beta$-equilibrated} NS matter predicted by the
stiff-type CDM3Y$n$ interactions.

The NS pressure obtained with different in-medium NN interactions are compared
with the empirical NS pressure deduced from the recent astronomical observation.
In general, the EOS given by the soft-type interactions tend to give pressure lower
the empirical values at high densities. In particular, the D1S version of the Gogny
interaction gives negative pressure at baryon densities $n_{\rm b} \gtrsim  2 n_0$
and violates, therefore, le Chatelier's principle that ensures the NS stability. The
adiabatic sound velocity estimated from the NS pressure given by two versions
of the soft-type M3Y-P$n$ interaction becomes superluminal at high baryon
densities and violates, therefore, the causality condition, and the EOS
has been corrected by hand in this case for further use in the TOV equations.
It seems, therefore, likely that there could be uncertainties in a well
parametrized effective (density dependent) NN interaction that are not visible
at low nuclear densities, and its success in the nuclear structure study is not
sufficient to validate its extrapolation to supranuclear densities.

Different EOS's of the NS core supplemented by the Sly4 EOS of the NS crust
given by the compressible liquid drop model have been used for the input
of the Tolman-Oppenheimer-Volkov equations to study how different behaviors of the
symmetry energy affect the model prediction of the NS properties. The EOS's
obtained with the stiff-type interactions were found to give consistently reasonable
description of the empirical data for the NS mass and radius, and to comply well with the
causality condition. In comparison with the same empirical NS data, the soft-type
interactions were found less successful, especially, the two versions of the famous
Gogny interaction certainly need an appropriate modification before they can be
used in the TOV equations to study structure of neutron star.

The vital role of the NM symmetry energy has been demonstrated in our specific
test of the CDM3Y$n$ interactions, where we found a significant reduction
of the maximum gravitational mass $M_{\rm G}$ and radius $R_{\rm G}$ away from
the empirical boundaries when the slope of the NM symmetry energy is changed from
the \emph{stiff} behavior to the \emph{soft} one. It is natural to expect that if
hyperons (and other hypothetical constituents like kaons or quark matter) are included
at high baryon densities, the $M_{\rm G}$ and $R_{\rm G}$values given by the
soft-type interactions could be driven to region lying well below all existing empirical
estimates.

\section*{Acknowledgments}
The present research has been
supported by the National Foundation for Scientific and Technological
Development (NAFOSTED) under Project Nr. 103.04.07.09. The first two authors
gratefully acknowledge the financial support from IPN Orsay and the LIA FVPPL
Programme for their short research stays at IPN Orsay in 2010. DTK thanks Betty
Tsang for her helpful comments and discussions.


\begin{thebibliography}{99}
\bibitem{Ba08} B.A. Li, L.W. Chen, and C.M. Ko, Phys. Rep.
{\bf 464}, 113 (2008).
\bibitem{Bet90} H.A. Bethe, Rev. Mod. Phys. {\bf 62}, 801 (1990).
\bibitem{Su94} K. Summiyoshi and H. Toki, Astrophys. J. {\bf 422}, 700 (1994).
\bibitem{Su95} K. Summiyoshi, K. Oyamatsu, and H. Toki, Nucl. Phys. {\bf A595}, 327 (1995).
\bibitem{Dou01} F. Douchin and P. Haensel,
 Astronomy \& Astrophys., {\bf 380}, 151 (2001).
\bibitem{Lat04} J.M. Lattimer and M. Prakash, Science {\bf 304}, 536 (2004);
 J.M. Lattimer and M. Prakash, Phys. Rep. {\bf 442}, 109 (2007).
\bibitem{Kl06} T. Kl\"ahn {\it et al.}, Phys. Rev. C {\bf 74}, 035802 (2006).
\bibitem{Bal07} M. Baldo and C. Maieron, J. Phys. G {\bf 34}, R243 (2007).
\bibitem{Be77} G. Bertsch, J. Borysowicz, H. McManus, and W.G. Love,
Nucl. Phys. {\bf A284}, 399 (1977).
\bibitem{An83} N. Anantaraman, H. Toki, and G.F. Bertsch,
 Nucl. Phys. {\bf A398}, 269 (1983).
\bibitem{Kho93} D.T. Khoa and W. von Oertzen, Phys. Lett. {\bf B304}, 8 (1993).
\bibitem{Kho95} D.T. Khoa and W. von Oertzen, Phys. Lett. {\bf B342}, 6 (1995).
\bibitem{Kho96} D.T. Khoa, W. von Oertzen, and A.A. Ogloblin, Nucl. Phys.
 {\bf A602}, 98 (1996).
\bibitem{Kho97} D.T. Khoa, G.R. Satchler, and W. von Oertzen,
 Phys. Rev. C {\bf 56} (1997) 954.
\bibitem{Kho07} D.T. Khoa, H.S. Than, and D.C. Cuong,
 Phys. Rev. C {\bf 76}, 014603 (2007).
\bibitem{Basu08} D.N. Basu, P. Roy Chowdlhury, C. Samanta, Nucl.
 Phys. {\bf A811}, 140 (2008); P. Roy Chowdlhury, C. Samanta, and D.N. Basu,
 Phys. Rev. C {\bf 80}, 011305(R) (2009).
\bibitem{Na02} H. Nakada and M. Sato, Nucl. Phys. {\bf A699}, 511 (2002).
\bibitem{Na03} H. Nakada, Phys. Rev. C {\bf 68}, 014316 (2003).
\bibitem{Na08} H. Nakada, Phys. Rev. C {\bf 78}, 054301 (2008); H. Nakada,
 Phys. Rev. C {\bf 82}, 029902 (2010).
\bibitem{Kho05} D.T. Khoa and H.S. Than, Phys. Rev. C {\bf 71}, 044601 (2005).
\bibitem{Kho07r} D.T. Khoa, W. von Oertzen, H.G. Bohlen, and S.
 Ohkubo, J. Phys. {\bf G34}, R111 (2007).
\bibitem{Kho09} N.D. Chien and D.T. Khoa, Phys. Rev. C {\bf 79}, 034314 (2009).
\bibitem{Tha09} H.S. Than, D.T. Khoa, and N.V. Giai,
 Phys. Rev. C {\bf 80}, 064312 (2009).
\bibitem{Be91} J.F. Berger, M. Girod, and D. Gogny,
 Comp. Phys. Comm. {\bf 63}, 365 (1991).
\bibitem{Ch08} F. Chappert, M. Girod, and S. Hilaire,
 Phys. Lett. B {\bf 668}, 420 (2008).
\bibitem{Ch98} E. Chabanat, P. Bonche, P. Haensel, J. Meyer, and R. Schaeffer,
 Nucl. Phys. {\bf A635}, 231 (1998).
\bibitem{Lat91} J.M. Lattimer, C.J. Pethick, M. Prakash, and P. Haensel,
 Phys. Rev. Lett. {\bf 66}, 2701 (1991).
\bibitem{Lat94} J.M. Lattimer, K.A. Van Riper, M. Prakash, and M. Prakash,
 Astrophys. J. {\bf 425}, 802 (1994).
\bibitem{Pa04} D. Page, J.M. Lattimer, M. Prakash, and A.W. Steiner,
 Astrophys. J. Suppl. Series {\bf 155}, 623 (2004).
\bibitem{Dou00} F. Douchin, P. Haensel, and J. Meyer,
  Nucl. Phys. {\bf A665}, 419 (2000).
\bibitem{Zuo99} W. Zuo, I. Bombaci, and U. Lombardo, Phys. Rev. C {\bf 60},
 024605 (1999).
\bibitem{Bra85} M. Brack, C. Guet, and H.B. H\aa kansson,
 Phys. Rep. {\bf 123}, 276 (1985).
\bibitem{Pea00} J.M. Pearson and R.C. Nayak, Nucl. Phys. {\bf A668}, 163 (2000).
\bibitem{Tsa09} M.B. Tsang, Y. Zhang, P. Danielewicz, M. Famiano, Z. Li,
  W.G. Lynch, and A.W. Steiner, Phys. Rev. Lett. {\bf 102}, 122701 (2009);
  M.B. Tsang, Z. Chajecki, D. Coupland, P. Danielewicz, F. Famiano, R. Hodges,
 M. Kilburn, F. Lu, W.G. Lynch, J. Winkelbauer, M. Youngs, and Y.X. Zhang,
 Prog. Part. Nucl. Phys. {\bf 66}, 400 (2011).
\bibitem{Da02} P. Danielewicz, R. Lacey and W.G. Lynch,
 Science {\bf 298}, 1592 (2002).
\bibitem{Ono03} A. Ono, P. Danielewicz, W.A. Friedman, W.G. Lynch,
 and M.B. Tsang, Phys. Rev. C {\bf 68}, 051601(R) (2003).
\bibitem{Ba09} Z. Xiao, B.A. Li, L.W. Chen, G.C. Yong, and M. Zhang,
 Phys. Rev. Lett. {\bf 102}, 062502 (2009).
\bibitem{She07} D.V. Shetty, S.J. Yennello, and G.A. Souliotis,
 Phys. Rev. C {\bf 76}, 024606 (2007).
\bibitem{Sh07} D.V. Shetty, S.J. Yennello, and G.A. Souliotis, Nucl.
 Inst. and Meth. in Phys. Res. B {\bf 261}, 990 (2007).
\bibitem{Ch05} L.W. Chen, C.M. Ko, and B.A. Li, Phys. Rev. Lett. {\bf 94},
 032701 (2005)
\bibitem{Da03} P. Danielewicz, Nucl. Phys. {\bf A727}, 233 (2003).
\bibitem{Aru04} P. Arumugam, B.K. Sharma, P.K. Sahu, S.K. Patra, Tapas Sil,
 M. Centelles, and X. Vi\~nas, Phys. Lett. B {\bf 601}, 51 (2004).
\bibitem{Tod05} B. G. Todd-Rutel and J. Piekarewicz,
 Phys. Rev. Lett. {\bf 95}, 122501 (2005).
\bibitem{Pie09} J. Piekarewicz and M. Centelles,
 Phys. Rev. C {\bf 79}, 054311 (2009).
\bibitem{Pie07} J. Piekarewicz, Phys. Rev. C {\bf 76}, 064310 (2007).
\bibitem{Cen09} M. Centelles, X. Roca-Maza, X. Vinas, and M. Warda,
 Phys. Rev. Lett. {\bf 102}, 122502 (2009).
\bibitem{Tri08} L. Trippa, G. Col\`o, and E. Vigezzi,
 Phys. Rev. C {\bf 77}, 061304(R) (2008).
\bibitem{Br00} B.A. Brown, Phys. Rev. Lett. {\bf 85}, 5296 (2000).
\bibitem{Fur02} R.J. Furnstahl, Nucl. Phys. {\bf A706}, 85 (2002).
\bibitem{Ak98} A. Akmal, V.R. Pandharipande, and D.G. Ravenhall,
 Phys. Rev. C {\bf 58}, 1804 (1998).
\bibitem{Gan10} S. Gandolfi, A.Yu. Illarionov, S. Fantoni, J.C. Miller, F. Pederiva,
 and K.E. Schmidt, Mon. Not. R. Astron. Soc. {\bf 404}, L35 (2010).
\bibitem{Ba05} V. Baran, M. Colonna, V. Greco, and M. Di Toro,
 Phys. Rep. {\bf 410}, 335 (2005).
\bibitem{Sto03} J.R. Stone, J.C. Miller, R. Koncewicz, P.D. Stevenson,
 and M.R. Strayer, Phys. Rev. C {\bf 68}, 034324 (2003).
\bibitem{Wir88} R.B. Wiringa, V. Fiks, and A. Fabrocini,
 Phys. Rev. C {\bf 38}, 1010 (1988).
\bibitem{DBHFa} E.N.E. van Dalen, C. Fuchs, and A. Faessler,
 Eur. Phys. J. A {\bf 31}, 29 (2007).
\bibitem{Li06} Z.H. Li, U. Lombardo, H.J. Schulze, W. Zuo, L.W. Chen,
 and H.R. Ma, Phys. Rev. C {\bf 74}, 047304 (2006).
\bibitem{Je77} J.P. Jeukenne, A. Lejeune, and C. Mahaux,
 Phys. Rev. C {\bf 16}, 80 (1977).
\bibitem{Ozel10} F.  \"{O}zel, G. Baym, and T. G\"{u}ver,
 Phys. Rev. D {\bf 82}, 101301(R) (2010).
\bibitem{Ste10} A.W. Steiner, J.M. Lattimer, and E.F. Brown,
 Astrophys. J. {\bf 722}, 33 (2010).
\bibitem{Bom01} I. Bombaci, {\it Isospin Physics in Heavy Ion Collisions at
Intermediate Energies}, Edited by B.A. Li and W.U.  Schr\"{o}der (Nova Science,
New York, 2001), p. 35.
\bibitem{Ioffe} http://www.ioffe.ru/astro//NSG/NSEOS/index.html.
\bibitem{She11} G. Shen, C.J. Horowitz, and S. Teige,
 Phys. Rev. C {\bf 83}, 035802 (2011).
\bibitem{Glen2} N.K. Glendenning, {\it Compact Stars: Nuclear Physics,
Particle Physics and General Relativity} (Springer: Springer-Verlag New York,
Inc. 2000).
\bibitem{Lat05} J.M. Lattimer and B.F. Schutz, Astrophys. J. {\bf 629}, 979 (2005).
\bibitem{Wor08} A. Worley, P.G. Krastev, and B.A. Li, Astrophys. J. {\bf 685}, 390 (2008).
\bibitem{Cot02} J. Cottam, F. Paerls, and M. Mendez, Nature {\bf 420}, 51 (2002).
\bibitem{Rho74} C.E. Rhoades, Jr., and R. Ruffini, Phys. Rev. Lett. {\bf 32}, 324 (1974).
\bibitem{Pod05} Ph. Podsiadlowski, J.D.M. Dewi, P. Lesaffre, J.C. Miller, W.G. Newton,
  and J.R. Stone, Mon. Not. R. Astron. Soc. {\bf 361}, 1243 (2005).

\end{thebibliography}
\end{document}